\documentclass[12pt,letterpaper]{article}
\usepackage{anysize,graphicx,amsfonts,amsmath,amssymb,hyperref,float,setspace}
\usepackage[font=small]{caption,subcaption}
\numberwithin{equation}{section}
\begin{document}
%========+=========+=========+=========+=========+=========+=========+=========+
\begin{titlepage}

\hbox to \hsize{\hspace*{0 cm}\hbox{\tt }\hss
    \hbox{\small{\tt }}}

\vskip1cm
\centerline{\Large\bf Electric hyperscaling violating solutions in }
\centerline{\Large\bf Einstein-Maxwell-dilaton gravity with $R^2$ corrections}

\vspace{1 cm}
 \centerline{\large Daniel K. O'Keeffe\footnote{dokeeffe@physics.utoronto.ca}\, and Amanda W. Peet\footnote{awpeet@physics.utoronto.ca}}

\vspace{0.3cm}
\begin{center}
{\it Department of Physics, \\{} University of Toronto, \\{} Toronto, Ontario, \\{} Canada M5S 1A7.}
\end{center}
 
\vspace{1 cm}

\begin{abstract}

In the context of holography applied to condensed matter physics, we study Einstein-Maxwell-dilaton theory with curvature squared corrections. This theory has three couplings $\eta_i$ for the three $R^2$ invariants and two theory functions: a dilaton potential $V(\phi)$ and a dilaton-dependent gauge coupling $f(\phi)$. We find hyperscaling violating solutions of this theory, parametrized by dynamical critical exponent $z$ and HSV parameter $\theta$. We obtain restrictions on the form of the theory functions required to support HSV-type solutions using three physical inputs: the null energy condition, causality $z\geq 1$, and $d_{\rm eff}\equiv d-\theta$ lying in the range $0<d_{\rm eff}\leq d$. The NEC constraints are linear in the $\eta_i$ and (quartic) polynomial in $d,z,\theta$. The allowed ranges of $z,\theta$ change depending on the signs of $\eta_i$. For the case of Einstein-Weyl gravity, we further narrow down the theory functions and solution parameters required for crossover solutions interpolating between HSV, $AdS_{d+2}$ near the boundary, and $AdS_2\times \mathbb{R}^d$ in the deep interior.  

\end{abstract}

\end{titlepage}

\begin{spacing}{0.9}
\tableofcontents
\end{spacing}

%========+=========+=========+=========+=========+=========+=========+=========+
\section{Introduction}\label{sec:intro}

The AdS/CFT correspondence is a remarkable construction which has sparked many new opportunities to study the detailed structure of strongly coupled quantum field theories. Among new avenues of investigation it has spawned are applications to modelling the quark-gluon plasma and condensed matter systems. The goal of modelling strongly coupled field theory systems at quantum critical points will be the context for the work reported here. Our perspective will be bottom-up, in the sense that we will seek particular classes of spacetime solutions in curvature squared gravity with dilaton potential and dilaton-dependent gauge couplings in order to seek out physical constraints on the theory functions and parameters and on parameters of solutions within it. Finding string theory embeddings for this class of models and analyzing technical stability properties (ghosts, etc) of the solutions that we investigate is beyond the scope of this work.  

Condensed matter systems typically do not possess relativistic symmetry. For instance, for field theories at finite charge density Lorentz invariance is broken by the presence of a current \cite{Hartnoll_Metallic}. Breaking of relativistic symmetry in the field theory implies that the bulk gravity/string dual should also break relativistic symmetry. Two major directions have been pursued in this context: spacetimes with Schr\"odinger symmetry \cite{son} and spacetimes with Lifshitz symmetry \cite{KLM}. Aspects of the dictionary are better developed for Lifshitz, such as holographic renormalization \cite{Ross_2011}\cite{Baggio_2012}, so we choose this as our context.

Lifshitz quantum critical points are invariant under the scaling symmetry
\begin{equation}{\label{scaling}}
t \rightarrow \lambda^{z} t \, , \qquad x_{i} \rightarrow \lambda x_{i} \, .
\end{equation}
where $z$ is the dynamical critical exponent. In \cite{KLM} a candidate gravity dual for Lifshitz fixed points was proposed, with spacetime metric
\begin{equation}{\label{lifintro}}
ds_{d+2}^{2} = L^{2}\left( -\mathcal{R}^{2 z} dt^{2} + \frac{d {\mathcal{R}}^{2}}{{\mathcal{R}}^{2}} + {\mathcal{R}}^{2} dx_{i}^{2}\right) \, ,
\end{equation}
where ${\mathcal{R}}$ is the radial coordinate, which ranges from ${\mathcal{R}} \rightarrow 0$ in the interior to ${\mathcal{R}} \rightarrow \infty$ at asymptopia.  The bulk spacetime respects Lifshitz scaling symmetry with ${\mathcal{R}} \rightarrow \lambda^{-1} {\mathcal{R}}$. Here, $d$ is the number of transverse dimensions $x_{i}$ and $L$ sets the length scale in the bulk.  

The Lifshitz metric and its finite temperature counterpart are exact solutions to Einstein gravity with a nontrivial matter sector. Two popular options for the matter sector are Einstein gravity coupled to a massive gauge field \cite{taylor},\cite{mcgreevy},\cite{hsvlif_2012} and Einstein-Maxwell-dilaton theory \cite{Tarrio_Vandoren},\cite{sachdevhsv},\cite{Charmousis}  
\begin{equation}{\label{einsteinmaxdil}}
S = {\frac{1}{16\pi G_N}}\int d^{d+2} x \sqrt{-g} \left ( R - \frac{1}{2} (\partial \phi)^{2} - V(\phi)  - \frac{1}{4}f(\phi)F^{2} \right) \, .
\end{equation}
In order to obtain Lifshitz solutions, it suffices to consider $f(\phi) \propto e^{\lambda_{1} \phi}$ and $V(\phi) = \Lambda$, where $\lambda_{1}$ is a constant and $\Lambda$ is the (negative) cosmological constant.  

Einstein-Maxwell-dilaton theory also supports a broader class of interesting spacetime geometries, the hyperscaling violating (HSV) metric   
\begin{equation}{\label{hsvintro}}
ds_{d+2}^{2} = L^{2} \left( - {\mathcal{R}}^{2(z-\theta/d)} dt^{2} + \frac{d{\mathcal{R}}^{2}}{{\mathcal{R}}^{2(1+\theta/d)}} + {\mathcal{R}}^{2(1-\theta/d)} dx_{i}^{2} \right) \, ,
\end{equation}
where $z$ is the dynamical critical exponent and $\theta$ is the hyperscaling violation parameter \cite{Gouteraux},\cite{Ugajin},\cite{Kachruhsv}. Metrics of HSV form are not scale invariant, but rather transform covariantly,
\begin{equation}
ds^{2}_{d + 2} \rightarrow \lambda^{2\theta/d} ds^2_{d + 2} \, .
\end{equation}
From the perspective of the dual theory, hyperscaling is the property that the free energy of the system should scale with its na{\"{\i}}ve dimension.  At finite temperature,  theories with hyperscaling have an entropy density which scales with temperature as $S \sim T^{d/z}$.  When hyperscaling is violated, there is a modified relationship, $S \sim T^{(d-\theta)/z}$, indicating the system lives in an effective dimension $d_{\rm{eff}} = (d-\theta)$ \cite{sachdevhsv},\cite{sachdevreview}. Candidate HSV gravity duals for systems of this sort will be the focus of this study.

Compressible phases of matter have strongly coupled quantum critical points in $2{}+1$ dimensions, making them obvious targets for holographic modelling. 
The HSV sub-case $\theta = d-1$ is particularly interesting because it describes the case of strange metals \cite{Hartnoll_Metallic}, a type of non-Fermi liquids.
In $D=d+2=4$ specifically, compressible non-Fermi liquid states are known from field theory analysis to have dynamical critical exponent $z=3/2$ up to three loop order and $\theta=1$ \cite{sachdevhsv}. 
Another motivation for studying candidate HSV gravity dual spacetimes in Einstein gravity is that there are \cite{sachdevhsv},\cite{sachdevreview} logarithmic violations of the area law for entanglement entropy.

We may ask which types of theory functions $f(\phi),V(\phi)$ can support HSV solutions and, if so, what physical parameters $z,\theta$ might be available.
For HSV solutions of Einstein-Maxwell-dilaton gravity, it suffices \cite{sachdevhsv} to take $f(\phi) \propto e^{\lambda_{1} \phi}$ and $V(\phi) \propto -e^{\lambda_{2} \phi}$, where $\lambda_{1}$ and $\lambda_{2}$ are constants and the dilaton runs logarithmically $\phi(\mathcal{R}) \propto \ln(\mathcal{R})+$const.
From the condensed matter perspective, having HSV solutions in Einstein gravity causes an embarrassment of riches, in the sense that it gives too many allowed values of $z,\theta$. A natural question from a microscopic perspective is whether or not introducing curvature squared corrections might help constrain the parameter space more tightly. 

Introducing curvature squared corrections to Einstein gravity alters the structure of the null energy condition (NEC), which we use as a primary tool to discriminate physical solutions from unphysical ones. Accordingly, via the NEC, solution parameters $z,\theta$ can be constrained in terms of theory parameters $\{\eta_i\}$, whose magnitude must be small in order that the semiclassical approximation we make in the gravity sector be believable. In addition to the NEC, we will insist on two basic requirements motivated from the field theory side: that the physical effective dimension $d_{\rm eff}$ for the dual field theory be positive \cite{sachdevhsv} and that $z\ge1$ to ensure causal signal propagation.

Curvature squared corrections are also motivated from study of singularities in Lifshitz-type and HSV-type solutions. First, consider Lifshitz.
At first glance, it appears that the Lifshitz gravity dual of \cite{KLM} is nonsingular, because all curvature invariants remain finite in the interior. However, the Lifshitz-type geometries display divergent tidal forces in the interior \cite{KLM} \cite{Mann_Copsey_2011}, \cite{Horowitz_Way_2011} which disturb string probes. 
For Lifshitz solutions with a magnetic ansatz for the gauge field, logarithmic running of the dilaton runs the gauge coupling to infinity as ${\mathcal{R}}\rightarrow 0$. In \cite{Kachru_Lif_2012}, it was shown for $D=4$ that quantum corrections to $f(\phi)$ can stabilize the dilaton and replace the deep interior geometry with $AdS_{2} \times \mathbb{R}^{2}$.
For electric Lifshitz solutions, the gauge field runs to weak coupling in the interior so quantum corrections to $f(\phi)$ do not provide a mechanism for resolving the tidal force singularity. However, recently it has been found \cite{lifcorrect} that in $D=4$ curvature squared corrections are capable of stabilizing the dilaton of electric Lifshitz, crossing over to $AdS_{2} \times \mathbb{R}^{2}$. We will build on this observation. Another feature of the electric equations of motion is that demanding that the spacetime be asymptotically Lifshitz (i.e. at ${\mathcal{R}}\rightarrow\infty$) makes the gauge coupling function formally diverge there. This was addressed in \cite{lifcorrect} which displayed a crossover to $AdS_{4}$ in the UV.  Lifshitz solutions in Einstein-Weyl were also studied in \cite{LifWeyl_2012}.
 
For HSV-type spacetimes with Einstein as the gravity sector, the situation is more involved. Tidal forces still generally diverge in the interior, but are avoided for specific ranges of $z$ and $\theta$ as pointed out in \cite{shaghoulian_2011}. Curvature invariants remain finite in the interior for \cite{Mann_Copsey_2012} $\theta > 0$, but diverge at the boundary, necessitating a UV completion to $AdS$ there. 
Magnetic HSV solutions display the same type of logarithmic running as their Lifshitz cousins and become strongly coupled in the interior. Using the same quantum corrections to $f(\phi)$ as  \cite{Kachru_Lif_2012}, \cite{HSV_ads2} constucted flows from HSV in to $AdS_{2} \times \mathbb{R}^{2}$ in the interior and out to $AdS_{4}$ at the boundary.
Electric HSV solutions, like for Lifshitz, do not run to strong coupling in the interior. Motivated by the observations in \cite{lifcorrect}, we will investigate  curvature squared solutions with hyperscaling violation and investigate whether there are IR and UV completions. 

The paper is organized as follows. In Section \ref{sec:r2hsv}, we look for HSV-type solutions to Einstein-Maxwell-dilaton theory with curvature squared corrections, and present the theory functions $f(\phi), V(\phi)$ needed to support these solutions. 
In Section \ref{sec:plots} we discuss how the null energy condition (NEC) along with the constraints $0<d_{\rm eff}\leq d$ and $z\geq1$ restricts polynomial combinations of solution parameters $z,\theta$ and theory parameters $\{\eta_i\}$.
In Section \ref{sec:crossover}, we discuss the question of crossovers between HSV, (a) $AdS_D$ asymptotically, and (b) $AdS_{2} \times \mathbb{R}^{d}$ in the deep interior, supported by the curvature squared corrections.
In Section \ref{sec:discussion}, we summarize our findings and comment on possible directions for future work.

%========+=========+=========+=========+=========+=========+=========+=========+
\section{Hyperscaling violation in Einstein-Maxwell-dilaton gravity with curvature squared corrections}{\label{sec:r2hsv}}

We will be interested in classes of models with curvature squared corrections to Einstein gravity, coupled to a $U(1)$ gauge field and a scalar which we will refer to as the dilaton. The action for the class of models we study is of the form
\begin{align}{\label{action}}
S = \frac{1}{16 \pi G_{N}} \displaystyle\int d^{d+2} x \sqrt{-g} &\left ( R  -\frac{1}{2} (\partial \phi)^{2} - V(\phi) -\frac{1}{4} f(\phi)F^{2} \right . \notag \\ &\left . + 
\eta_{1} R_{\mu \nu \rho \sigma} R^{\mu \nu \rho \sigma} + \eta_{2} R_{\mu \nu} R^{\mu \nu} + \eta_{3} R^{2}   \right .  \bigg{)} \,.
\end{align}
Here, the $\eta_{i}$ are constant couplings for the higher curvature terms measured in units of $\ell_P$.  

Our goal in this section is to find HSV-type solutions to the field equations of this model. We will use the metric ansatz
\begin{equation}{\label{hsvmet}}
ds_{d+2}^{2} = -L^{2} r^{2 \alpha} dt^{2} + L^{2} \frac{dr^{2}}{r^{2 \beta}} + L^{2} r^{2} dx_{i}^{2} \,.
\end{equation}
This form of the ansatz is chosen in order that the (fourth order) equations of motion have a good chance of being tractable analytically. Here, $d$ is the number of {transverse} dimensions, $D=d+2$ is the bulk spacetime dimension, and $L$ sets the overall length scale. The dynamical exponent $z$ and the hyperscaling violation parameter $\theta$ are related to the parameters $\alpha$ and $\beta$ by
\begin{equation}
\alpha = \frac{dz -\theta}{d-\theta} \,, \qquad \beta = \frac{d}{d-\theta} \,.
\end{equation}
This metric (\ref{hsvmet}) is related to that of the previous section (\ref{hsvintro}) by a coordinate transformation $\mathcal{R} = r^{d/(d-\theta)}$ which will help make our equations simpler.  The Riemann curvature components of (\ref{hsvmet}) are
\begin{align}
&R^{t r}_{~~t r} = -{\alpha(\alpha+\beta-1)} r^{2(\beta-1)}/L^2\,,  & R^{r i}_{~~r i} = -{\beta} r^{2(\beta-1)}/{L^{2}} \,, \notag \\
&R^{i t}_{~~t i} = {\alpha}r^{2(\beta - 1)}/L^2\,,  & R^{i j}_{~~j i} =r^{2(\beta-1)} /L^2\,,
\end{align}
where repeated indices $i,j$ are not summed over.  From this it is straightforward to obtain the Ricci tensor and Ricci scalar for the equations of motion.

In order to support a HSV spacetime, it will be necessary to include a nontrivial potential for the the dilaton $V(\phi)$.  In the case of pure Lifshitz ($\theta = 0$ or $\beta = 1$), the dilaton potential reduces to a constant and plays the role of a cosmological constant: $V(\phi) \rightarrow \Lambda_{\rm Lif}$ as $\theta \rightarrow 0$.   
Later on when we investigate the possibility of producing $AdS$ completions to the HSV geometries in both the UV and IR, we will see that $V(\phi)$ will also set the individual $AdS$ scales. That is, we will look for a mechanism by which the higher curvature corrections to the action stabilize the dilaton at some constant value $\phi = \phi_{0}$.  When evaluated on this solution, $V(\phi_{0})$ will set the $AdS$ scale for us.  

The lore for the hyperscaling violating metrics of the form (\ref{hsvmet}) is that the minimum ingredients needed to support such a metric are a gauge coupling $f(\phi)$ and dilaton potential $V(\phi)$ that are exponentials in the dilaton $\phi$ \cite{sachdevhsv}\cite{Kachruhsv}.  This is valid in the limit of matter plus Einstein gravity, but not in the case when higher curvature terms like those in our action (\ref{action}) are present, as was pointed out recently in \cite{NewShagh}.  Indeed, as we will see, the form of the dilaton potential and gauge coupling will need to be modified in order to support the HSV spacetime in our theory with curvature squared corrections. We will see that not all parameters  support HSV solutions, and we will explore the admissible ranges of  $\{\eta_i\}$ using two tools: (1) the null energy condition (NEC) and (2) constraints on parameters from the condensed matter side. The hope is that bottom-up investigations of this sort may help serve as a partial guide to top-down string embedders. We now turn to the equations of motion and solving them.

The Maxwell field equation takes the form
\begin{equation}{\label{Maxwell}}
\nabla_{\mu} \left[ f(\phi) F^{\mu \nu} \right] = 0 \,,
\end{equation}
while the dilaton equation of motion is
\begin{equation}{\label{dilaton}}
\Box \phi - \partial_{\phi} f(\phi) F_{\mu \nu} F^{\mu \nu} - \partial_{\phi} V(\phi) = 0 \,.
\end{equation}
Via repeated application of the Bianchi identities, the equations of motion for the metric become
\begin{align}{\label{einstein}}
T_{\mu \nu} \equiv \widetilde{G}_{\mu \nu} &= R_{\mu \nu} - \frac{1}{2} g_{\mu \nu} R + 2 \eta_{1} R_{\mu \lambda \rho \sigma} R_{\nu}^{~\lambda \rho \sigma} + (4\eta_{1} + 2 \eta_{2}) R_{\mu \lambda \nu \sigma} R^{\lambda \sigma} - 4 \eta_{1} R_{\mu \lambda} R_{\nu}^{~\lambda} \notag \\
&-(2\eta_{1} + \eta_{2} + 2\eta_{3}) \nabla_{\mu} \nabla_{\nu} R + (4 \eta_{1} + \eta_{2}) \Box R_{\mu \nu} + 2 \eta_{3} R R_{\mu \nu} \notag \\
&-\frac{1}{2} g_{\mu \nu} \left [\eta_{1} R_{\alpha \beta \rho \sigma}R^{\alpha \beta \rho \sigma} + \eta_{2} R_{\lambda \sigma} R^{\lambda \sigma} + \eta_{3} R^{2} - (\eta_{2} + 4 \eta_{3}) \Box R \right ] \,,
\end{align}
while the energy-momentum tensor is
\begin{equation}{\label{enmom}}
T_{\mu \nu} = \frac{1}{2} (\partial_{\mu} \phi) (\partial_{\nu} \phi) -\frac{1}{2} g_{\mu \nu} V(\phi) - \frac{1}{4} g_{\mu \nu} (\partial \phi)^{2} + \frac{1}{2} f(\phi) \left (F_{\mu \sigma} F_{\nu}^{~\sigma} - \frac{1}{4} g_{\mu \nu} F_{\lambda \sigma} F^{\lambda \sigma} \right ) \,.
\end{equation}

Making use of an electric ansatz for the gauge field yields a solution to the Maxwell field equation (\ref{Maxwell})
\begin{equation}
F^{r t} = \frac{Q}{\sqrt{-g} f(\phi)} = 
\frac{Q}{f(\phi) L^{d+2}r^{\alpha-\beta+d}}  \,,
\end{equation}
where $Q$ is a constant of integration.  

Note that for a magnetic HSV solution in a radial ansatz we would seek $F_{(2)}={\cal{B}}(r)dx\wedge dy$, where 
\begin{equation}
{\mathcal{B}}(r) = {\frac{P}{f(\phi) L^{d-4} r^{\alpha-\beta+d-4}}} \,.
\end{equation}
We will stick with the electric case. 

Now, using the components of the energy-momentum tensor (\ref{enmom}), the equations for the metric may be recast in a more useful form,
\begin{align}{\label{einequiv1}}
&V(\phi(r)) = -\frac{1}{L^{2}} \left ( r^{-2} \widetilde{G}_{i i} + r^{2\beta} \widetilde{G}_{r r} \right ) \,, \notag \\
&(\partial \phi)^{2} = \frac{2}{L^{2}} \left ( r^{-2\alpha} \widetilde{G}_{t t} + r^{2\beta} \widetilde{G}_{r r} \right ) \,,\notag \\
&\frac{Q^{2}}{f(\phi(r)) L^{2d}} r^{-2d} = \frac{1}{2L^{2}} \left ( r^{-2\alpha} \widetilde{G}_{t t} + r^{-2} \widetilde{G}_{i i} \right ) \,,
\end{align}
where there is no sum on repeated indices $i$.
It is straightforward to obtain the components of $\widetilde{G}_{\mu \nu}$, and they turn out to be a sum of two competing powers of $r$. This is easiest to see by raising one index:
\begin{align}\label{eom}
&\widetilde{G}^{t}_{~t} = -\frac{C_{1}}{L^{2}} r^{2(\beta-1)} - \frac{C_{2}}{L^{2}} r^{4(\beta-1)} \,, \notag \\
&\widetilde{G}^{r}_{~ r} = \frac{C_{3}}{L^{2}} r^{2(\beta-1)} + \frac{C_{4}}{L^{2}} r^{4(\beta -1)} \,,\notag \\
&\widetilde{G}^{i}_{~ i} = \frac{C_{5}}{L^{2}} r^{2 (\beta-1)} + \frac{C_{6}}{L^{2}} r^{4 (\beta -1)} \,.
\end{align}
Here, the $C_{i}$ are constants in $\alpha$, $\beta$, $d$, $\eta_{1}$, $\eta_{2}$, and $\eta_{3}$. The details of the long expressions are relegated to the Appendix; let us briefly summarize their features. 
First, the odd constants.  $C_{1}(d,\beta)$ is linear in $\beta$ and quadratic in $d$; $C_{3}(d,\alpha)$ is linear in $\alpha$ and quadratic in $d$; and $C_{5}(d,\alpha,\beta)$ is linear in $\beta$ and quadratic in $d$ and $\alpha$.  
Second, the even constants.  
 $C_{2}$, $C_{4}$ and $C_{6}$ are linear in $\{\eta_i\}$, quartic in $d$ (with coefficients depending on $\{\eta_i\}$), quartic in $\alpha$, and cubic in $\beta$ (except for $C_4$ which is quadratic). Only $C_{2}$, $C_{4}$ and $C_{6}$ contain information about the higher curvature terms in the action (\ref{action}), so those are the ones to watch.  

The final result for the field equations in this ansatz simplifies to
\begin{align} 
&V(\phi(r)) = -\frac{1}{L^{2}} \left [ D_{1} r^{2(\beta-1)} + D_{2} r^{4(\beta -1)} \right ] \label{einequiv2a} \,, \\
&(\partial \phi)^{2} = \frac{2}{L^{2}} \left [ D_{3} r^{2(\beta-1)} + D_{4} r^{4(\beta-1)} \right ] \label{einequiv2b} \,, \\
&\frac{Q^{2}}{f(\phi(r)) L^{2 d}} r^{-2 d} = \frac{1}{2 L^{2}} \left [ D_{5} r^{2(\beta-1)} + D_{6} r^{4(\beta-1)} \right ] \,, \label{einequiv2c}
\end{align}
where the constants $\left \{ D_{1} \ldots D_{6} \right \}$ are linear combinations of the $\left \{ C_{1} \ldots C_{6} \right \}$ constants as follows,
\begin{align}\label{Dee4Dee6}
&D_{1}(d,\alpha,\beta) = C_{5} + C_{3}, \quad D_{2}(d,\alpha,\beta,\eta_{1},\eta_{2},\eta_{3}) = C_{4} + C_{6} \,, \\
&D_{3}(d,\alpha,\beta) = C_{1} + C_{3}, \quad D_{4}(d,\alpha,\beta,\eta_{1},\eta_{2},\eta_{3}) = C_{2} + C_{4} \,, \\
&D_{5}(d,\alpha,\beta) = C_{1} + C_{5}, \quad D_{6}(d,\alpha,\beta,\eta_{1},\eta_{2},\eta_{3}) = C_{2} + C_{6}  \,.
\end{align}
Once we integrate (\ref{einequiv2b}), we have both the required form of the theory functions and the form of the solutions, from (\ref{einequiv2a}-\ref{einequiv2c}). Note that we believe the curvature squared HSV solutions presented here to be novel in the context of AdS/condensed matter but unlikely to be so as GR spacetimes.

Before proceeding to analytic solutions, we should ask what kind of restrictions we can impose on the space of parameters of the theory.  A very natural choice is to insist on satisfying the null energy condition $T_{\mu \nu} N^{\mu} N^{\nu} \ge 0$ in order to insure that we are dealing with a sensible matter source for the model.  Here, the inequality must hold for any arbitrary null vector $N^{\mu}$.  Using the field equations (\ref{einstein}) for the metric, this statement may be translated into the condition $\widetilde{G}_{\mu \nu} N^{\mu} N^{\nu} \ge 0$.  
An appropriate null vector is
\begin{equation}
N^{t} = \left ( \displaystyle\sum_{i = 1}^{d} s_{i}^{2} + s_{r}^{2} \right ) \frac{1}{L r^{\alpha}} \,,  \qquad
N^{r} = s_{r} \frac{r^{\beta}}{L} \,, \qquad
N^{i} = s_{i} \frac{1}{L r} \,,
\end{equation}
where $s_{r}$ and $s_{i}$ ($d$ of them) are arbitrary positive constants.  Using this $N^\mu$, the NEC translates into the following conditions on the constants $D_{i}$:
\begin{align}
&D_{3}(d,\alpha,\beta) \ge 0 \,, \qquad D_{4}(d,\alpha,\beta,\eta_{1},\eta_{2},\eta_{3}) \ge 0 \,,\\
&D_{5}(d,\alpha,\beta) \ge 0 \,, \qquad D_{6}(d,\alpha,\beta,\eta_{1},\eta_{2},\eta_{3}) \ge 0 \,.
\end{align}
(Note that there are no conditions on $D_1$ or $D_2$ coming from the NEC.)
Two of these conditions, $D_{3} \ge 0$ and $D_{5} \ge 0$, collapse into the simple relations
\begin{align}
&(z-1)(z-\theta+d) \ge  0 \label{NEC1} \,, \\
&(d-\theta)(d(z-1)-\theta) \ge 0 \label{NEC2} \,,
\end{align}
respectively.  These are identical to conditions found when applying the NEC to HSV solutions of Einstein-Maxwell-dilaton theory \cite{sachdevhsv} \cite{Kachruhsv}.  This had to be the case, as our higher curvature model contains the Einstein gravity terms.  Only the conditions $D_{4} \ge 0$ and $D_{6} \ge 0$ depend on the couplings $\eta_{i}$.

Now let us move to solving these equations analytically. We may start by solving the differential equation for $(\partial \phi(r))^{2}$, (\ref{einequiv2b}) directly, to get
\begin{align}{\label{dilatonSol}}
\phi(r) = &- \frac{\sqrt{2 D_{3}}}{\beta - 1} \left [ \sqrt{1 + \frac{D_{4}}{D_{3}} r^{2(\beta-1)}} - \text{arccsch}\left (\sqrt{\frac{D_{4}}{D_{3}}} r^{\beta-1} \right ) \right ] \notag \\
&+ \frac{\sqrt{2 D_{3}}}{\beta -1} \left [ \sqrt{1 + \frac{D_{4}}{D_{3}}} - \text{arccsch}\left ( \sqrt{\frac{D_{4}}{D_{3}}} \right ) \right ] + c \,,
\end{align}
where $c$ is a constant.  Note that, at first glance, this solution may seem to be undefined in the limit that $\beta \rightarrow 1$ ($\theta \rightarrow 0 \Rightarrow \alpha = z$). However, this is just an illusion.  The constant ($r$-independent) terms in $\phi(r)$ are precisely those needed to cancel the divergence from the first two terms, and the limit is well defined:
\begin{equation}
\phi(r) \rightarrow \left. -\sqrt{2D_{3} + 2D_{4}}\;\right|_{\beta \rightarrow 1} \ln(r)  + c \,.
\end{equation}
This is precisely the kind of logarithmic behaviour of the dilaton we would expect for a purely Lifshitz behaviour \cite{lifcorrect}.  

It is instructive to study the asymptotic behaviour of $\phi(r)$.  By expanding the solution (\ref{dilatonSol}) as $r \rightarrow \infty$ and $r \rightarrow 0$, we can see what the dilaton is doing in the UV and IR respectively.  The result in the UV is
\begin{equation}{\label{phiUV}}
\phi(r) \vert_{r\rightarrow \infty} \rightarrow - \frac{\sqrt{2 D_{4}}}{(\beta - 1)} \,r^{\beta - 1} + c \,,
\end{equation}
while in the IR it is
\begin{equation}{\label{phiIR}}
\phi(r) \vert_{r\rightarrow 0} \rightarrow -\frac{\sqrt{2 D_{3}}}{(\beta - 1)} \ln \left ( \frac{1}{2} \sqrt{\frac{D_{4}}{D_{3}}} r^{\beta - 1} \right ) + c \,.
\end{equation}
In a putative string theory embedding, this would imply that the string coupling involving $e^\phi$ is diverging in the deep interior where we know the null singularity lurks, and dies out to zero out at the boundary.

In order to satisfy the remaining gravity equations, we need a form for $V(\phi)$ and $f(\phi)$ such that
\begin{align}
&V(\phi(r)) = -\frac{1}{L^{2}} \left [D_{1} r^{2(\beta-1)} + D_{2} r^{4(\beta-1)} \right ] \,,
&f(\phi(r)) = \frac{2 Q^{2}}{L^{2(d-1)}} \frac{r^{-2(\beta + d-1)}}{\left ( D_{5} + D_{6} r^{2(\beta-1)} \right )} \,.
\end{align}
In general, for arbitrary $D_{i}$ (i.e., arbitrary $\eta_i$ and arbitrary $\alpha,\beta$), it is difficult to  invert the solution (\ref{dilatonSol}) for $\phi(r)$. The analytic functions encountered are Lambert W-functions, which do not have  visually pleasant representations, so we do not display them here. Instead, we leave $V(\phi(r))$ and $f(\phi(r))$ in implicit form along with (\ref{dilatonSol}) describing $\phi(r)$ or alternately (\ref{einequiv2b})) describing $d\phi(r)/dr$. 

Regardless of the detailed form of $f(\phi)$ and $V(\phi)$, it is straightforward to examine their asymptotic behaviours using the results in (\ref{phiUV}) and (\ref{phiIR}):
\begin{align}
& V(\phi)\vert_{r\rightarrow\infty} \rightarrow -\frac{(\beta-1)^{2}}{2 D_{4} L^{2}} \left [ D_{1} + \frac{(\beta-1)^{2}}{2 D_{4}} D_{2} (\phi - c)^{2} \right ] (\phi - c)^{2} \,, \\
&V(\phi)\vert_{r\rightarrow 0} \rightarrow -\frac{4 D_{3}}{D_{4} L^{2}} \exp \left ( \frac{2 (\beta - 1)}{\sqrt{2 D_{3}}} (c - \phi) \right )  \left [ D_{1} + \frac{4 D_{3} D_{2}}{D_{4}} \exp \left ( \frac{2 (\beta - 1)}{\sqrt{2 D_{3}}} (c - \phi) \right ) \right ] \,. \notag
\end{align}
Obviously, these formul\ae\ are not valid for $\eta_W\rightarrow 0$; there the form changes back to what we expect from an Einstein gravity sector. Note that the magnitude of the potential is controlled by $1/\eta_W$, and that at large $r$ the potential naturally measures $\phi$ in units of $\sqrt{\eta_W}$.

Recall that $f(\phi)$ plays the role of the coupling for the Maxwell field: $f(\phi(r)) \sim 1/g_M^{2}$.  Hence $g_M \sim (L^{d-1})/(\sqrt{2} Q) \left ( D_{5} +D_{6} r^{2(\beta-1)} \right )^{1/2} r^{2(\beta +d -1)}$.  In terms of $z$ and $\theta$: $2(\beta-1) = (2 \theta)/(d-\theta)$ and $\beta +d -1 = d +\theta/(d-\theta)$.  Hence, $g \rightarrow 0$ as $r \rightarrow 0$ provided that $\theta \ge 0$ and $(d-\theta) > 0$, meaning that we get to weak gauge coupling in the interior of the spacetime for physically sensible parameter ranges, as desired.  Furthermore,
\begin{equation}
f(\phi) \vert_{r\rightarrow\infty} \rightarrow  \frac{2 D_{4} Q^{2}}{(\beta-1)^{2} L^{2(d-1)}} \frac{\left [ -\frac{(\beta-1)}{\sqrt{2 D_{4}}} (\phi-c) \right ] ^{(-2d)/(\beta-1)}}{\left [ D_{5}  + \frac{D_{6} (\beta-1)^{2}}{2 D_{4}}(\phi-c)^{2}\right ](\phi-c)^{2}} \,.
\end{equation} 
As $r\rightarrow 0$, the coupling $f(\phi)$ goes to zero.
The remaining dilaton equation of motion (\ref{dilaton}) then collapses to
\begin{align}
&(\alpha + \beta + d -1) D_{3} - (\beta + d -1) D_{5} + (\beta - 1) D_{1} = 0 \label{dil1} \,, \\
&(\alpha + 2 \beta + d -2) D_{4} - (d+2\beta-2) D_{6} + 2(\beta-1) D_{2} = 0 \label{dil2} \,.
\end{align}
It is easy to verify that these two equations are satisfied identically for all $d$, $\alpha$, $\beta$ and $\eta_{i}$  by virtue of the ansatz.

Finally, we note that by saturating one of the NEC inequalities, $z=1+\theta/d$, it is possible to  reduce the complexity of $V(\phi),f(\phi)$ to power laws in the dilaton.

The next step is to explore which parameter ranges are physically admissible when we have HSV solutions in our theory with curvature squared corrections. Our main physics tool for investigating this will be the NEC, the details of which we derived earlier in this section. We now turn to visualizing the NEC constraints graphically.

%========+=========+=========+=========+=========+=========+=========+=========+
\section{Exploring parameter ranges using the NEC}\label{sec:plots}
In the previous section, we used $\{\eta_1,\eta_2,\eta_3\}$ to parametrize the curvature squared corrections to Einstein gravity in our model. It is convenient at this point to change basis to the more traditional basis $\{\eta_W,\eta_{GB},\eta_R\}$ where
\begin{equation}{\label{model}}
\mathcal{L}_{HC} = \eta_{W} C_{\mu \nu \lambda \sigma} C^{\mu \nu \lambda \sigma} + \eta_{GB} G + \eta_{R} R^{2} \,,
\end{equation}
where $G = R_{\mu \nu \rho \sigma} R^{\mu \nu \rho \sigma} - 4 R_{\mu \nu} R^{\mu \nu} + R^{2}$ is the usual Gauss-Bonnet term and the Weyl tensor is
\begin{equation}
C_{\mu \nu \rho \sigma} = R_{\mu \nu \rho \sigma} - \frac{2}{d} \left ( g_{\mu [ \rho} R_{\sigma ] \nu} - g_{\nu [ \rho} R_{\sigma ] \mu} \right ) + \frac{2}{d(d+1)} R\, g_{\mu [ \rho} g_{\sigma ] \nu} \,.
\end{equation}
Here, anti-symmetrization of the indices is defined as $T_{[\mu \nu]} = \frac{1}{2} \left ( T_{\mu \nu} - T_{\nu \mu} \right )$.
It is straightforward to work out the relation between the coupling constants above and those in our original basis:
\begin{align*}
&\eta_{1} = \eta_{W} + \eta_{GB}\,, \\
&\eta_{2} = -\frac{4}{d} \eta_{W} - 4 \eta_{GB}\,, \\
&\eta_{3} = \frac{2}{d(d+1)} \eta_{W} + \eta_{GB} + \eta_{R}\,.
\end{align*}
There a few sanity checks that we can make in this basis.  In particular, the Gauss-Bonnet term should vanish for $d=1$ and is topological for $d=2$ ($D = d+2 = 3$ and $D = d+2 = 4$), hence the equations of motion should be independent of $\eta_{GB}$ for $d\le2$.  This is straightforwardly verified.  Furthermore, the Weyl tensor vanishes in AdS, so in the limit that $\alpha = 1$ and $\beta = 1$ ($z=1$ and $\theta = 0$, respectively), we expect the equations of motion to be independent of $\eta_{W}$.  Again, this is straightforwardly verified. In fact, it is true for $\alpha = 1$ even for $\beta \ne 1$ ($z=1$, $\theta \ne 0$, respectively).   

It is difficult to visualize the influence of the three independent couplings $\eta_{GB},\eta_W,\eta_R$ at once. We will address this complexity in stages by examining the cases (i) only one parameter turned on; (ii) two parameters turned on; and (iii) three parameters turned on. Again, our main physics tool will be the NEC.

%========+=========+=========+=========+=========+=========+=========+=========+
\subsection{Gauss-Bonnet gravity}\label{sec:GB}

Gauss-Bonnet gravity is the case where $\eta_{R} = \eta_{W} = 0$. We begin with this case because it turns out to be the simplest one.

In terms of $z$ and $\theta$, the NEC for HSV solutions reduces down to
\begin{align}
&-\eta_{GB} \,d\, (d-1)(d-2)\,[d(z-1)-\theta] \ge 0 \label{nec_etagb_1} \,, \\
&-\eta_{GB} \,d\,(d-1)(d-2)(z-1)\,[d^{2}+dz-d\theta+2\theta] \ge 0 \label{nec_etagb_2}  \,.
\end{align}
Notice that both conditions (\ref{nec_etagb_1}) and (\ref{nec_etagb_2}) are trivial when $d=1$ or $d=2$. As we pointed out in the previous subsection $(3.1)$, the Gauss-Bonnet term vanishes for $d=1$ and is topological for $d=2$ (i.e. $D = d+2 = 4$) and so does not contribute to the equations of motion. This is why the NEC reduces to (\ref{NEC1}) and (\ref{NEC2}) of Einstein gravity.  

For $d>2$ where the Gauss-Bonnet term is nontrivial, combining (\ref{NEC1}) and (\ref{NEC2}) with (\ref{nec_etagb_1}) and (\ref{nec_etagb_2}) produces different outcomes depending on the sign of $\eta_{GB}$. 
For $\eta_{GB} > 0$, there is only one way to support HSV: $z=1$ and $\theta = 0$, i.e. the $AdS_{d+2}$ limit.  For $\eta_{GB} \le 0$, we end up with the Einstein gravity NEC. 

\begin{figure}[H]
\centering
	\begin{subfigure}{0.49\linewidth} \centering
		\includegraphics[scale=0.425]{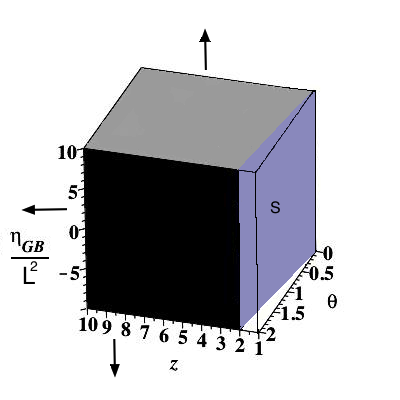}
		\caption{$d=2$, only $\eta_{GB} \neq 0$.} \label{fig:etaGB_d2}
	\end{subfigure}
	\begin{subfigure}{0.49\linewidth} \centering
		\includegraphics[scale=0.425]{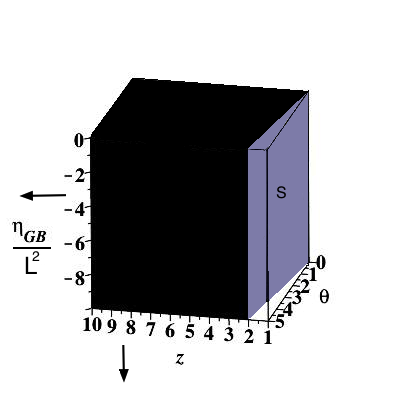}
		\caption{$d=5$, only $\eta_{GB} \neq 0$.} \label{fig:etaGB_d5}
	\end{subfigure}
\caption{Restrictions on $\eta_{GB}$ from the NEC for $d=2$ and $d=5$ respectively. 
Note that there is only one arrow towards decreasing $\eta_{GB}$ for the $d=5$ case (\ref{fig:etaGB_d5}).} \label{fig:etaGB_full}
\end{figure}

In general terms, we want to understand how the NEC conditions in our $R^2$ HSV model restrict the theory parameters $\eta_{i}$ and the solution parameters $z,\theta$. To see how, it is instructive to plot the inequalities as a function of $\eta_{i}$, $z$, and $\theta$ while fixing the number of transverse dimensions $d$. For the Gauss-Bonnet case, the permissible parameter regions are shown in Fig.\ref{fig:etaGB_full}, for two values of $d$. The plots look so simple here because the constraints are linear. For every other case that we will discuss in this section, the constraints will look more opaque and we use the plots to help shed light on them. 

Our greyscale conventions for figure features are as follows. (i) Allowed regions are bounded by the metallic grey surface(s) labeled ``S'', which we will refer to in the following as the constraint surface.  (ii) An arrow indicates that the object shown -- whose cross section is depicted as a light grey surface perpendicular to the base of the arrow -- continues semi-infinitely in the direction indicated by that arrow. (iii) Black indicates that either (a) the HSV parameter $\theta$ leaves the physically acceptable regime of $0 \le \theta < d$ or (b) the $\eta_i$ parameters become inadmissible, i.e. do not support HSV solutions satisfying the NEC. 

\vspace{-12pt}
%======FIGURE======%
%======FIGURE======%
\vspace{-12pt}

%========+=========+=========+=========+=========+=========+=========+=========+
\subsection{Einstein-Weyl gravity}\label{sec:W}

Einstein-Weyl gravity is what we get when we set $\eta_{GB} = \eta_{R} = 0$. 
This is of interest in its own right because the Weyl tensor vanishes in $AdS$. 

In this case, we find that
\begin{align}
& -\frac{4 \eta_{W}}{(d+1)L^{2}} (d-1)(d+2\beta-2)(\alpha-1)(\alpha+\beta-1)(\alpha-3\beta-d+1) \geq 0 \,,{\label{D4EW}}  \\
& -\frac{4 \eta_{W}}{(d+1)L^{2}} (d-1)(\alpha-1)(\alpha+\beta-1)(3\beta+\alpha+d-3)(d\alpha-2d\beta+2-2\beta-d^{2}) \geq 0 \,,{\label{D6EW}}
\end{align}

Before we move to the plot, let us verify that our solution recovers the known Lifshitz solution \cite{lifcorrect} in the limit that $\beta \rightarrow 1$ (which implies that $\alpha \rightarrow z$).  This is indeed precisely what we obtain: $V(\phi(r))$ reduces to the cosmological constant for the Lifshitz case, $\phi(r) \propto \ln(r)$ and $f(\phi(r))$ is an exponential of the dilaton.  

A curious sub-case is the one with $\alpha = 1$ (so $z=1$), but $\beta \ne 1$, that is, the ``purely" hyperscaling violating solution.  In this case, we find a logarithmic dilaton and an exponential potential
\begin{equation}
\phi(r) = -\sqrt{2d(1-\beta)} \ln(r) +{\rm const} \,,
\end{equation}
\begin{equation}
V(\phi) = -\frac{\tilde{A}}{L^{2}} \exp \left ( \sqrt{\frac{2(1-\beta)}{d}} \phi \right ) \,,
\end{equation}
where $\tilde{A}$ is a constant which depends on $d$ and $\beta$, and 
\begin{equation}
f(\phi) \rightarrow \infty \,.
\end{equation}
The logarithmic running of the dilaton and a potential that is exponential in $\phi$ is to be expected here \cite{sachdevhsv} \cite{lifcorrect}.
The fact that $f(\phi) \rightarrow \infty$ is not physically disturbing.  Recall the solution to Maxwell's equations (\ref{Maxwell}) is 
$F^{r t} = Q/[\sqrt{-g} f(\phi)] = Qr^{\beta - \alpha -d}/[f(\phi) L^{d+2}]$.  
Hence, as $f(\phi) \rightarrow \infty$, the field strength vanishes, meaning that the gauge field reduces to a constant.  This is to be expected as the role of the gauge field was to break the usual relativistic scaling symmetry to the non-relativistic Lifshitz case.  When $z=1$, this scaling symmetry is restored and the gauge field is no longer necessary.  

The NEC conditions in this case are
\begin{align}
&\eta_{W} [d(d-\theta)-d(z-2)+2\theta] \ge 0 \label{nec_etaw_1} \,,\\
&\eta_{W}[d^{3}-d^{2}(z+\theta-2)+(d+2)\theta] \ge 0 \label{nec_etaw_2} \,,
\end{align}
which we plot along with the other two NEC constraints (\ref{NEC1}) and (\ref{NEC2}).  Several example plots are shown below.

\vspace{-12pt}
%======FIGURE======%
\begin{figure}[H]
\centering
	\begin{subfigure}{0.49\linewidth} \centering
		\includegraphics[scale=0.425]{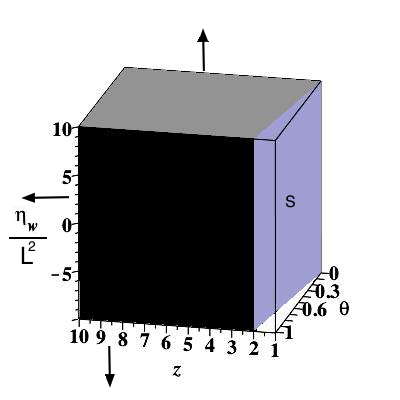}
		\caption{$d=1$, only $\eta_{W} \neq 0$} \label{fig:etaW_d1}
	\end{subfigure}
\caption{NEC restrictions on $\eta_{W}$ for $d=1$. 
} \label{fig:etaW_d1_full}
\end{figure}
%======FIGURE======%
\vspace{-12pt}

\vspace{-12pt}
%======FIGURE======%
\begin{figure}[H]
\centering
	\begin{subfigure}{0.49\linewidth} \centering
		\includegraphics[scale=0.425]{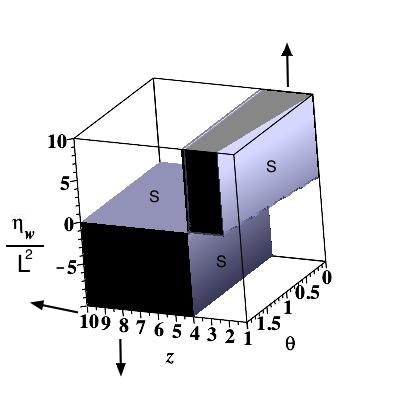}
		\caption{$d=2$, only $\eta_{W} \neq 0$.} \label{fig:etaW_d2_b}
	\end{subfigure}
	\begin{subfigure}{0.49\linewidth} \centering
		\includegraphics[scale=0.425]{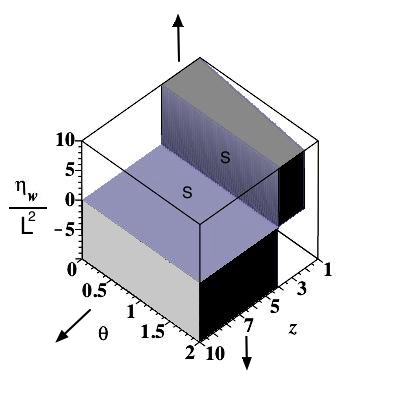}
		\caption{$d=2$, only $\eta_{W} \neq 0$, side view.} \label{fig:etaW_d2}
	\end{subfigure}
\caption{Restrictions on $\eta_{W}$ from the NEC for $d=2$. 
For $z<4$, the NEC is satisfied in the hexahedral region in the upper right hand side of both sub-figures. 
For $z>4$, the NEC is satisfied in the rectangular region in the lower left hand side of both sub-figures.
} \label{fig:etaW_d2_full}
\end{figure} 
%======FIGURE======%
\vspace{-12pt}

Figure (\ref{fig:etaW_d1_full}) shows the restrictions imposed by the NEC in $d=1$.  Notice that there are no constraints on $\eta_{W}$.  This is not troubling as in $d=1$, the conditions (\ref{nec_etaw_1}) and (\ref{nec_etaw_2}) vanish and what we are plotting then is nothing more than the conditions (\ref{NEC1}) and (\ref{NEC2}) familiar from Einstein gravity.  In $d=1$, the Weyl tensor vanishes for the HSV metric and so $\eta_{W}$ plays no role.

Figure (\ref{fig:etaW_d2_full}) shows the allowed regions for $d=2$.  Curiously, there is transition at $z=4$; for $z<4$ the NEC is satisfied in the hexahedral region in the upper right hand side, as seen in figures (\ref{fig:etaW_d2}) and (\ref{fig:etaW_d2_b}), which restricts $\eta_{W} \ge 0$.  For $z>4$, the situation is flipped and the allowed region is the box in the lower left hand side, restricting $\eta_{W} \le 0$.  When $z=4$ and $d=2$, the conditions (\ref{nec_etaw_1}) and (\ref{nec_etaw_2}) vanish and the NEC is satisfied for all sensible values of $\theta$ and there are no restrictions on $\eta_{W}$.  This is indicated in (\ref{nec_etaw_1}) and (\ref{nec_etaw_2}) by the plane cutting through the figures at $z=4$.  Curiously enough, for $d=2$ and $z=4$, the Weyl tensor does not vanish as it does for $d=1$ and for  $z=1$, and so is still contributing to the equations of motion.  
For $d>2$, a qualitatively similar transition in behaviour occurs, however the crossover now happens for a range of values of $z$ and $\theta$.  

From the perspective of condensed matter theory, there is interest in holographic theories with $d=2$, $\theta = d-1= 1$ and $z=3/2$, which are proposed to capture some of the mysterious physics of strange metal phases \cite{sachdevhsv}.  Figures (\ref{fig:etaW_d2}) and (\ref{fig:etaW_d2_b}) shows that this range sits comfortably within the hexahedral region in the upper right hand side.

%========+=========+=========+=========+=========+=========+=========+=========+
\subsection{{\texorpdfstring{$R^{2}$}{R2}} gravity}\label{sec:R}

\begin{figure}[H]
\centering
	\begin{subfigure}{0.49\linewidth} \centering
		\includegraphics[scale=0.425]{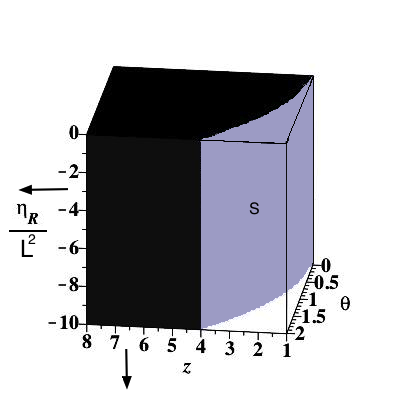}
		\caption{$d=2$, only $\eta_{R} \neq 0$.} \label{fig:etaR_d2}
	\end{subfigure}
	\begin{subfigure}{0.49\linewidth} \centering
		\includegraphics[scale=0.38]{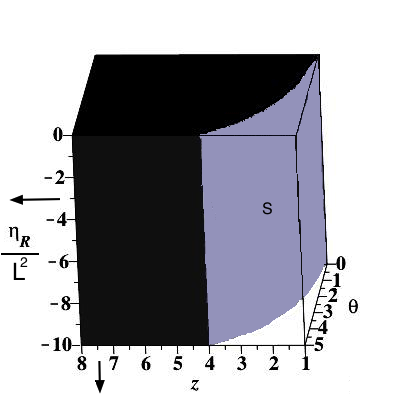}
			\caption{$d=5$, only $\eta_{R} \neq 0$.} \label{fig:etaR_d5}
	\end{subfigure}
\caption{Restrictions on $\eta_{R}$ from the NEC for $d=2$ and $d=5$.  
Note that there are two black sides in both plots, unlike earlier cases.} \label{fig:etaR_d1_d2_full}
\end{figure}

Consider the case that $\eta_{R} \neq 0$, $\eta_{GB} = 0$, and $\eta_{W} = 0$, so that $R^{2}$ is the only higher curvature contribution.  In this case, the NEC conditions are
\begin{equation}{\label{D4R}}
-\eta_{R} [(d+1) \theta^{2} + d^{2}(d+1) + 2(d+z)(dz-d\theta-\theta)][(d-8)\theta^{2} - d(d-2)z\theta + d^{3}(z-1)] \ge 0\,,
\end{equation}
\begin{equation}{\label{D6R}}
-\eta_{R} [(d+1) \theta^{2} + d^{2}(d+1) + 2(d+z)(dz-d\theta-\theta)][(z-1)(d(z-\theta+d+2\theta)] \ge 0\,,
\end{equation}
along with (\ref{NEC1}) and (\ref{NEC2}).  These conditions are shown in Fig.\ref{fig:etaR_d1_d2_full} for two representative values of $d$.

\vspace{-12pt}
%======FIGURE======%
%======FIGURE======%
\vspace{-12pt}

The results for $d>5$ are all qualitatively similar.  In all cases, the allowed region is bounded by the curved surface (``S" in figures (\ref{fig:etaR_d1_d2_full})). This surface is always bounded by lines along the $\eta_{R}$ axis at $z=0$ and $z=4$.  The NEC also restricts $\eta_{R} \le 0$ in all $d$.  In fact, the only way to support $\eta_{R} > 0$ is to set $z=1$ and $\theta = 0$.  In this case, both conditions (\ref{D4R}) and (\ref{D6R}) vanish.  

We can also consider turning on more than one coupling at a time.  In the next four sections we will investigate the constraints on multiple $\eta$s imposed by $d$, $z$ and $\theta$.

%========+=========+=========+=========+=========+=========+=========+=========+
\subsection{{\texorpdfstring{$R^{2}$}{R2}} and Weyl terms}\label{sec:RW}

\begin{figure}[H]
\centering
	\begin{subfigure}{0.3\linewidth} \centering
		\includegraphics[scale=0.425]{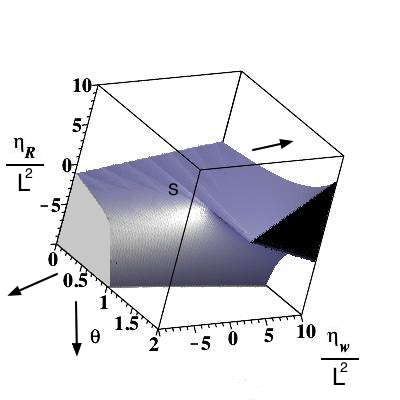}
		\caption{$d=2$, $z=2$} \label{fig:etaW_etaR_d2_z2}
	\end{subfigure}
	\hskip0.03\linewidth
	\begin{subfigure}{0.3\linewidth} \centering
		\includegraphics[scale=0.425]{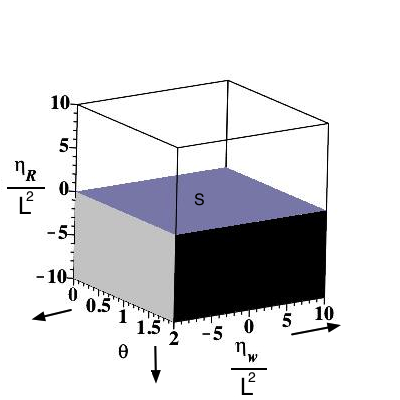}
		\caption{$d=2$, $z=4$} \label{fig:etaW_etaR_d2_z4}
	\end{subfigure}
	\hskip0.03\linewidth
		\begin{subfigure}{0.3\linewidth} \centering
		\includegraphics[scale=0.425]{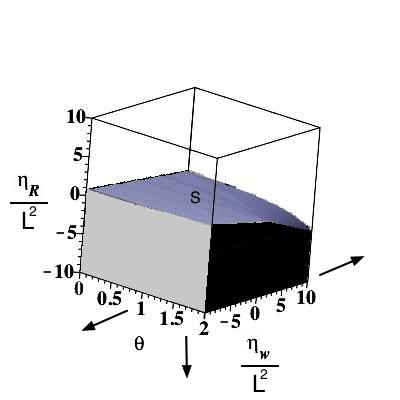}
		\caption{$d=2$, $z=6$} \label{fig:etaW_etaR_d2_z6}
	\end{subfigure}
\caption{Restrictions on $\eta_{W}$ and $\eta_{R}$ from the NEC for $d=2$ and several values of $z$. Notice the sharp change in behaviour in Fig.\ref{fig:etaW_etaR_d2_z4} where $d=2$, $z=4$.} \label{fig:etaW_etaR_d2}
\end{figure}

As a first example,  consider the case $\eta_{R} \neq 0$, $\eta_{W} \neq 0$, but $\eta_{GB} = 0$.  The conditions are
\begin{align}{\label{D4EWR}}
&-\eta_{R} \left \{ d (d+1) (\theta^{2} + d^{2}) + 2(d+z)(dz-d\theta-\theta)[(d-8)\theta^{2} - d(d-2)z\theta + d^{3}(z-1)] \right \} \notag \\
&+\frac{2 \eta_{W}}{(d+1)}d^{2} (d-1)(z-1)[d^{2} - (d-2) \theta][d(d-\theta) - d(z-2) + 2 \theta] \ge 0 \,,
\end{align}
\vspace{-12pt}
\begin{align}{\label{D6EWR}}
&-\eta_{R} \left \{d^{2} (z-1)(\theta^{2} + d^{2}) + 2(d+z)(dz-d\theta-\theta)(d(z-\theta+d) + 2\theta) \right \} \notag \\
&+\frac{2 \eta_{W}}{d(d+1)} d^{2} z(z-1)(d-1)[d^{3}-d^{2}(z+\theta-2)+(d+2)][dz + 2 \theta + d(d-\theta)] \ge 0 \,. 
\end{align}
Given a value for $d$, and $z$, it is instructive to plot these inequalities (along with (\ref{NEC1}) and (\ref{NEC2})) for the physically sensible range of $\theta$, this is depicted in Fig.\ref{fig:etaW_etaR_d2} for several different values.  
Note that because we now have two theory parameters varying in the plots, we have to fix one of the other parameters per plot to fit the plot into 3D. For clarity, we choose to fix $z$ for any given plot (as well as $d$, as before) in order to visualize the constraint surfaces for $\theta$.

\vspace{-12pt}
%======FIGURE======%
%======FIGURE======%
\vspace{-12pt}

The figures for $d=3,4,5$ look qualitatively similar, with only minor quantitative differences.

Notice that for $d=2$, there is an interesting change in behaviour at $z=4$ in Fig.\ref{fig:etaW_etaR_d2}.  At this point, all the contribution from the Weyl term vanishes from the NEC and we are left simply with the the conditions of pure $R^{2}$ gravity.  As we found in section (\ref{sec:R}), when only the $\eta_{R}$ term is non-zero, then physically sensible ranges of $z$ and $\theta$ restrict $\eta_{R} < 0$.  We see precisely this kind of behaviour in the case of having both $\eta_{R}$ and $\eta_{W}$ turned on; at $d=2$ and $z=4$, the $\eta_{W}$ contribution vanishes and we are left with only $\eta_{R}$ which is required to be less than or equal to zero, consistent with our previous result.   

$d=2$ and $z=4$ is special in that it is the unique combination of parameters for which the $\eta_{W}$ term does not contribute to the NEC.  At the level of the equations of motion (\ref{eom}), the relevant equations reduce down to those of pure $R^{2}$ gravity for this choice of parameters.  In dimensions other than $d=2$, a qualitatively similar transition is observed, but the transition is not as sharp as the contribution from the $\eta_{W}$ term never drops out completely.
 
%========+=========+=========+=========+=========+=========+=========+=========+
 \subsection{{\texorpdfstring{$R^{2}$}{R2}}  and Gauss-Bonnet terms}\label{sec:RGB}
 
\begin{figure}[H]
\centering
	\begin{subfigure}{0.3\linewidth} \centering
		\includegraphics[scale=0.425]{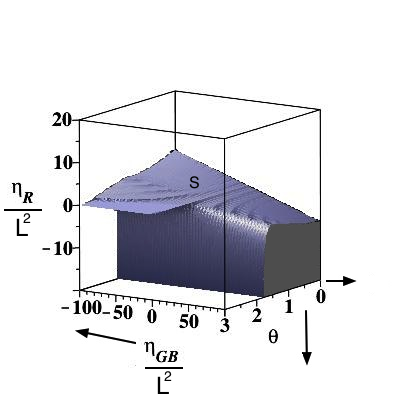}
		\caption{$d=3$, $z=2$} \label{fig:etaR_etaGB_d3_z2}
	\end{subfigure}
	\hskip0.03\linewidth
	\begin{subfigure}{0.3\linewidth} \centering
		\includegraphics[scale=0.425]{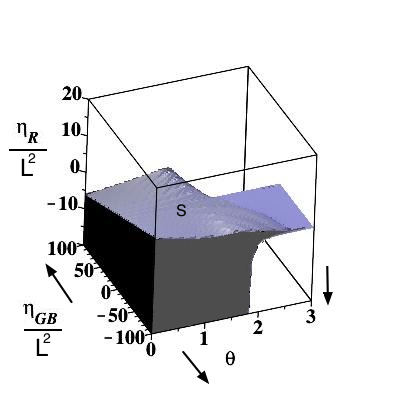}
		\caption{$d=3$, $z=2$, side view} \label{fig:etaR_etaGB_d3_z2_a}
	\end{subfigure}
	\hskip0.03\linewidth
	\begin{subfigure}{0.3\linewidth} \centering
		\includegraphics[scale=0.425]{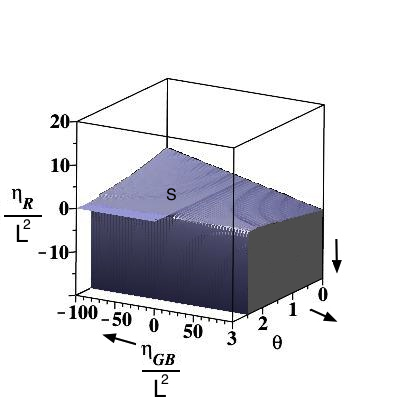}
		\caption{$d=3$, $z=3$} \label{fig:etaR_etaGB_d3_z3}
	\end{subfigure}
\caption{Restrictions on $\eta_{GB}$ and $\eta_{R}$ from the NEC for $d=3$ and $z=2$ and $z=3$.} \label{fig:etaR_etaGB_d3_a}
\end{figure}

\begin{figure}[H]
\centering
	\begin{subfigure}{0.49\linewidth} \centering
		\includegraphics[scale=0.425]{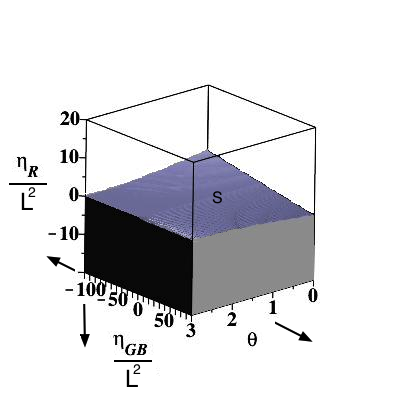}
		\caption{$d=3$, $z=4$} \label{fig:etaR_etaGB_d3_z4}
	\end{subfigure}
	\begin{subfigure}{0.49\linewidth} \centering
		\includegraphics[scale=0.425]{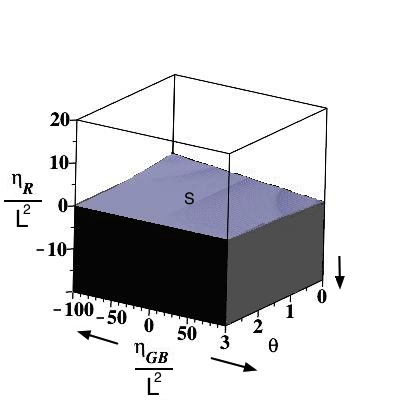}
		\caption{$d=3$, $z=6$} \label{fig:etaR_etaGB_d3_z6}
	\end{subfigure}
\caption{Restrictions on $\eta_{GB}$ and $\eta_{R}$ from the NEC for $d=3$ and $z=4$ and $z=6$. Notice the sharp change in behaviour compared (\ref{fig:etaR_etaGB_d3_a}) where $d=3$, $z<4$.} \label{fig:etaR_etaGB_d3_b}
\end{figure}

 Consider $\eta_{W} = 0$, $\eta_{GB} \neq 0$ and $\eta_{R} \neq 0$.  The NEC conditions are
 \begin{align}{\label{D4GBR}}
 &-\eta_{R} [ 2dz(d+z)+d^{2}(d+1)-2(d+1)z\theta-2d(d+1)+(d+1)\theta^{2}][(d-8)\theta^{2} \notag \\
 &-d(d-2)z \theta +d^{3}(z-1)] -\eta_{GB} (d-1)(d-2)(d-\theta)^{3}[d(z-1)-\theta] \ge 0 \,,
 \end{align}
 \begin{align}{\label{D6GBR}}
 &-\eta_{R} d(z-1)[d(z-\theta+d)+2\theta)][2dz(d+z)+d^{2}(d+1)-2(d+1)z\theta-2d(d+1)\notag \\
 &+(d+1)\theta^{2}] -\eta_{GB}(d-1)(d-2)(z-1)(d-\theta)^{2}[d(z-\theta+d)+2\theta] \ge 0 \,,
 \end{align}
which are to be supplemented by (\ref{NEC1}) and (\ref{NEC2}).  Notice that for $d=1$ and $d=2$, the Gauss-Bonnet contribution to (\ref{D4GBR}) and (\ref{D6GBR}) vanishes and we are left simply with pure $R^{2}$ gravity as in section (\ref{sec:R}).  Fig.\ref{fig:etaR_etaGB_d3_a} and Fig.\ref{fig:etaR_etaGB_d3_b} plot the restrictions on $\eta_{GB}$ and $\eta_{R}$ for $d=3$ and a few values of $z$.

\vspace{-12pt}
%======FIGURE======%
%======FIGURE======%
\vspace{-12pt}

\vspace{-12pt}
%======FIGURE======%	
%======FIGURE======%
\vspace{-12pt}

Notice the sharp change in behaviour for $z\ge4$ in Fig.\ref{fig:etaR_etaGB_d3_b}.    Below $z=4$, both positive and negative values of $\eta_{R}$ and $\eta_{GB}$ are allowed up to a maximum value of $\theta$ (for $z=2$, this value is $\theta=1.8$, for example).  Above this value of only $\eta_{GB} <0$ is allowed and $\eta_{R}$ is also severely restricted, as seen in Fig.\ref{fig:etaR_etaGB_d3_z2_a}.  In fact, this transition in behaviour is independent of $d$ and always occurs at $z=4$.  As in previous sections, $z=4$ turns out to be special.

\begin{figure}[H]
\centering
	\begin{subfigure}{0.49\linewidth} \centering
		\includegraphics[scale=0.425]{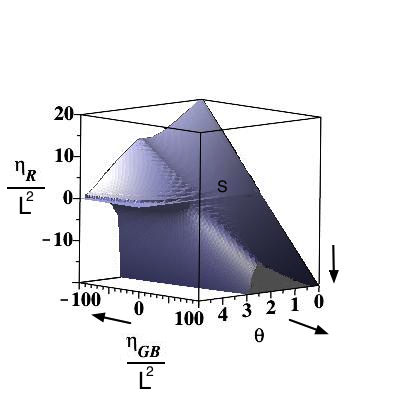}
		\caption{$d=5$, $z=2$} \label{fig:etaR_etaGB_d5_z2}
	\end{subfigure}
	\begin{subfigure}{0.49\linewidth} \centering
		\includegraphics[scale=0.425]{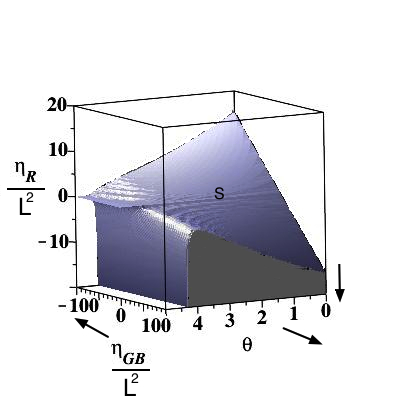}
		\caption{$d=5$, $z=3$} \label{fig:etaR_etaGB_d5_z3}
	\end{subfigure}
\caption{Restrictions on $\eta_{GB}$ and $\eta_{R}$ from the NEC for $d=5$ and $z=2$ and $z=3$.} \label{fig:etaR_etaGB_d5_a}
\end{figure}

\begin{figure}[H]
\centering
	\begin{subfigure}{0.49\linewidth} \centering
		\includegraphics[scale=0.425]{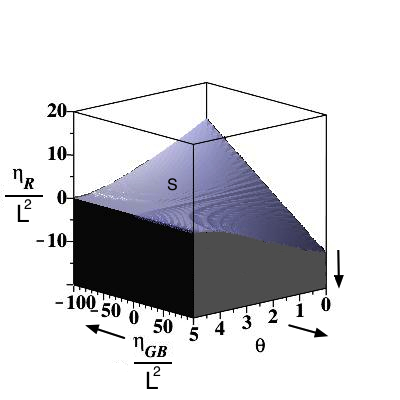}
		\caption{$d=5$, $z=4$} \label{fig:etaR_etaGB_d5_z4}
	\end{subfigure}
	\begin{subfigure}{0.49\linewidth} \centering
		\includegraphics[scale=0.425]{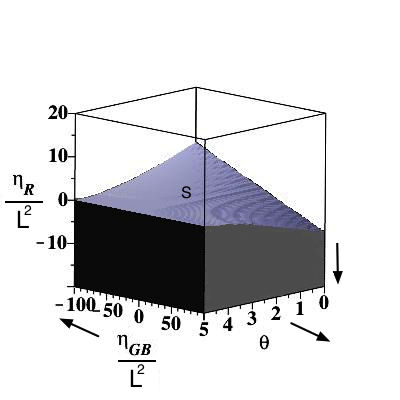}
		\caption{$d=5$, $z=6$} \label{fig:etaR_etaGB_d5_z6}
	\end{subfigure}
\caption{Restrictions on $\eta_{GB}$ and $\eta_{R}$ from the NEC for $d=5$ and $z=4$ and $z=6$. Notice the sharp change in behaviour compared to Fig.\ref{fig:etaR_etaGB_d5_a} where $d=5$, $z<4$.} \label{fig:etaR_etaGB_d5_b}
\end{figure}

For $d>3$, qualitatively similar behaviour is observed.  
%Fig.\ref{fig:etaR_etaGB_d4_a}, 
Fig.\ref{fig:etaR_etaGB_d5_a} and Fig.\ref{fig:etaR_etaGB_d5_b} provide a few salient examples.  

\vspace{-12pt}
%======FIGURE======%
%======FIGURE======%
\vspace{-12pt}

\vspace{-12pt}
%======FIGURE======%
%======FIGURE======%
\vspace{-12pt}

%========+=========+=========+=========+=========+=========+=========+=========+
 \subsection{Gauss-Bonnet and Weyl}\label{sec:GBW}
 
 Turning our attention to the case of $\eta_{R} = 0$ and $\eta_{GB} \neq 0$ and, $\eta_{W} \neq 0$, the conditions are
\begin{align}{\label{D4GBW}}
&2dz (z-1)(d-1)(d(d-\theta)+d(z-2)+2\theta) \eta_{W} \notag \\
& - (d-1)(d-2)(d+1)(d-\theta)^{3}(d(z-1)-\theta) \eta_{GB} \ge 0 \,,
\end{align}
\begin{align}{\label{D6GBW}}
&2z(z-1)(d-1)(d(d-\theta)+dz+2\theta)(d^{2}(d-\theta)-d^{2}(z-2)+(d+2)\theta)\eta_{W} \notag \\
&- (z-1)(d-1)(d-2)(d+1)(d-\theta)^{2}(d(z-\theta + d)+2\theta) \eta_{GB}  \ge 0 \,,
 \end{align}
which are to be supplemented by (\ref{NEC1}) and (\ref{NEC2}).  As we have seen in previous cases, for $d=1$ and $d=2$, the Gauss-Bonnet term does not contribute and we are back to simply the case of $\eta_{W} \neq 0$ examined in section (\ref{sec:W}).  Plots of allowed regions of $\eta_{GB}$ and $\eta_{W}$ are shown below for representative values of $d \ge 3$, $z$ and $\theta$.  
 
\vspace{-12pt}
%======FIGURE======%
\begin{figure}[H]
\centering
	\begin{subfigure}{0.3\linewidth} \centering
		\includegraphics[scale=0.425]{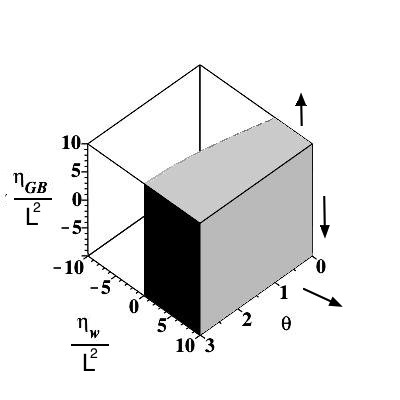}
		\caption{$d=3$, $z=2$.} \label{fig:etaW_etaGB_d3_z2_B}
	\end{subfigure}
	\hskip0.03\linewidth
	\begin{subfigure}{0.3\linewidth} \centering
		\includegraphics[scale=0.425]{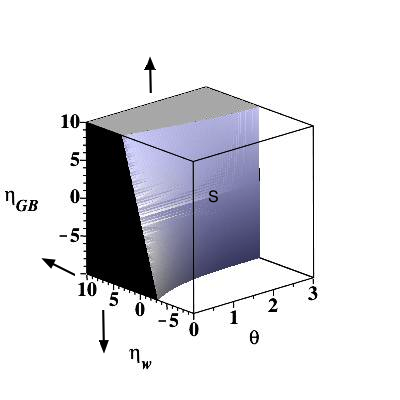}
		\caption{$d=3$, $z=2$, back view.} \label{fig:etaW_etaGB_d3_z2_A}
	\end{subfigure}
	\hskip0.03\linewidth
		\begin{subfigure}{0.3\linewidth} \centering
		\includegraphics[scale=0.425]{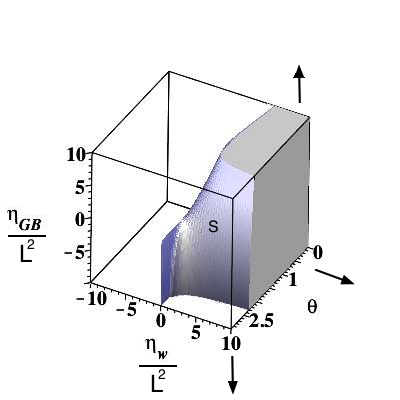}
		\caption{$d=3$, $z=4$.} \label{fig:etaW_etaGB_d3_z4_A}
	\end{subfigure}
	\begin{subfigure}{0.49\linewidth} \centering
		\includegraphics[scale=0.425]{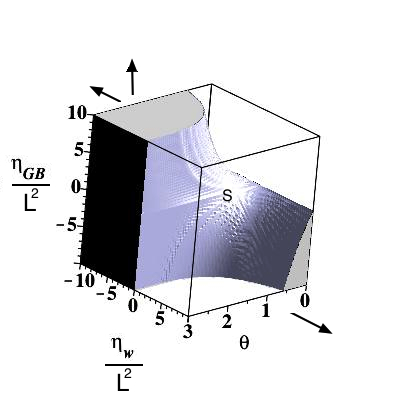}
		\caption{$d=3$, $z=5$.} \label{fig:etaW_etaGB_d3_z5_A}
	\end{subfigure}
	\begin{subfigure}{0.49\linewidth} \centering
		\includegraphics[scale=0.425]{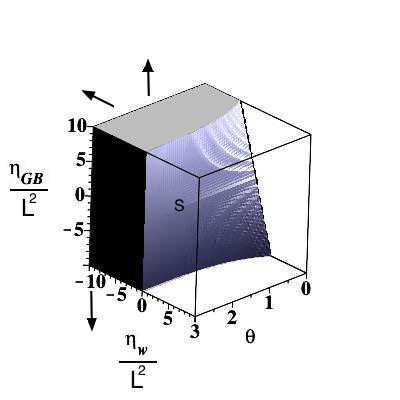}
		\caption{$d=3$, $z=6$.} \label{fig:etaW_etaGB_d3_z6_A}
	\end{subfigure}
\caption{Restrictions on $\eta_{GB}$ and $\eta_{W}$ from the NEC for $d=3$ and $z=2$, $z=4$, $z=5$ and $z=6$.}{\label{fig:etaW_etaGB_d3_full}}
\end{figure}
%======FIGURE======%
\vspace{-12pt}

For $d=3$ depicted above Fig.\ref{fig:etaW_etaGB_d3_full}, there are distinct transitions in behaviour that occur at $z=4$ and $z=5$.  Once we hit $z=6$, the allowed region is similar to that of $z=2$, except that the allowed value of $\eta_{W}$ are reflected by a minus sign.  This effect happens for higher dimensions as well, although the precise value of $z$ depends on $d$ and is not universal.  Following the common theme that we have seen in previous sections, values of $z$ around 4 mark a noticeable change in the allowed parameter region.  Higher dimensions display analogous behaviour.  We provide a few examples below for $d=4$ and $d=5$.

\vspace{-12pt}
%======FIGURE======%
\begin{figure}[H]
\centering
	\begin{subfigure}{0.3\linewidth} \centering
		\includegraphics[scale=0.425]{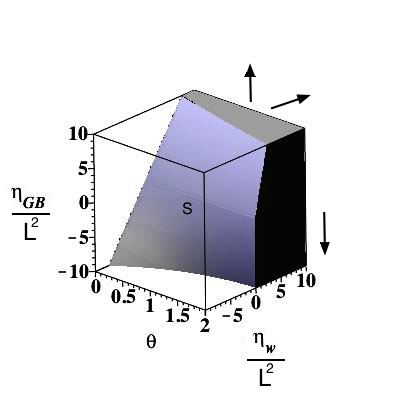}
		\caption{$d=4$, $z=\frac{3}{2}$.} \label{fig:etaW_etaGB_d4_z32_A}
	\end{subfigure}
	\hskip0.03\linewidth
	\begin{subfigure}{0.3\linewidth} \centering
		\includegraphics[scale=0.425]{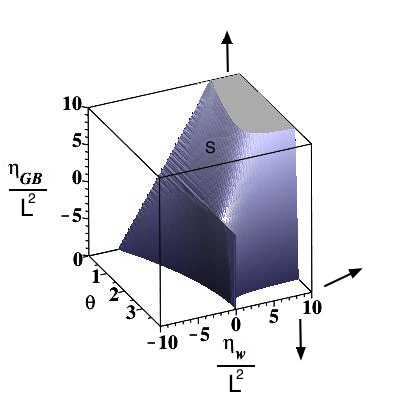}
		\caption{$d=4$, $z=4$.} \label{fig:etaW_etaGB_d4_z4_B}
	\end{subfigure}
	\hskip0.03\linewidth
	\begin{subfigure}{0.3\linewidth} \centering
		\includegraphics[scale=0.45]{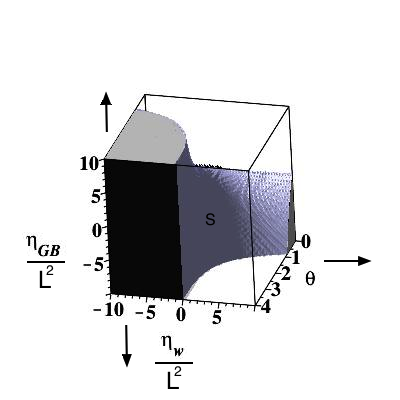}
		\caption{$d=4$, $z=6$.} \label{fig:etaW_etaGB_d4_z6_B}
	\end{subfigure}
\caption{Restrictions on $\eta_{GB}$ and $\eta_{W}$ from the NEC for $d=4$ and $z=\frac{3}{2}$, $z=4$ and $z=6$.}
\end{figure}
%======FIGURE======%
\vspace{-12pt}

\vspace{-12pt}
%======FIGURE======%
 \begin{figure}[H]
\centering
	\begin{subfigure}{0.3\linewidth} \centering
		\includegraphics[scale=0.425]{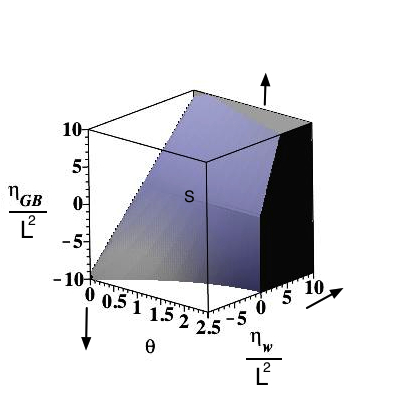}
		\caption{$d=5$, $z=\frac{3}{2}$.} \label{fig:etaW_etaGB_d5_z32_A}
	\end{subfigure}
	\hskip0.03\linewidth
	\begin{subfigure}{0.3\linewidth} \centering
		\includegraphics[scale=0.425]{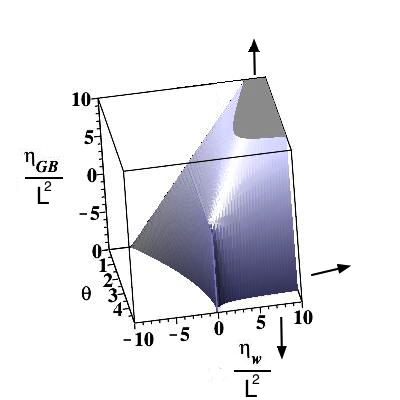}
		\caption{$d=5$, $z=4$.} \label{fig:etaW_etaGB_d5_z4_A}
	\end{subfigure}
	\hskip0.03\linewidth
	\begin{subfigure}{0.3\linewidth} \centering
		\includegraphics[scale=0.425]{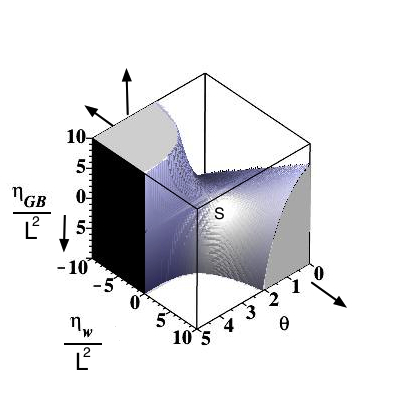}
		\caption{$d=5$, $z=6$.} \label{fig:etaW_etaGB_d5_z6_A}
	\end{subfigure}
\caption{Restrictions on $\eta_{GB}$ and $\eta_{W}$ from the NEC for $d=5$ and $z=\frac{3}{2}$, $z=4$ and $z=6$.}{\label{fig:etaW_etaGB_d5_full}}
\end{figure}
%======FIGURE======%
\vspace{-12pt}
 
%========+=========+=========+=========+=========+=========+=========+=========+
 \subsection{{\texorpdfstring{$R^{2}$}{R2}}, Gauss-Bonnet and Weyl}\label{sec:RGBW}
 
Finally, let us consider the case of having all three $\eta_{R}$, $\eta_{GB}$ and $\eta_{W}$ turned on at once. It is easiest to visualize the impact of different theory parameter choices by having  all three vary in our plots. This requires us to show different $z,\theta$ via different plots. So in this subsection, we are showing different slicings through the five-dimensional parameter space $\{\eta_{GB},\eta_W, \eta_R, z,\theta\}$ than we did in the single-$\eta$ and pair-of-$\eta$s cases. The plots may look less structured but this is just a slicing artefact.

\begin{figure}[H]
\centering
	\begin{subfigure}{0.49\linewidth} \centering
		\includegraphics[scale=0.425]{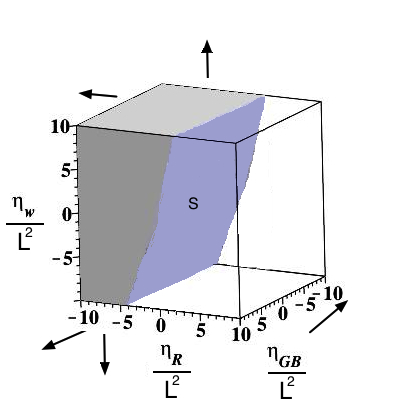}
		\caption{$d=3$, $z=2$, $\theta = 1$.} \label{fig:etaGB_etaR_etaW_d3_z2_theta1}
	\end{subfigure}
	\begin{subfigure}{0.49\linewidth} \centering
		\includegraphics[scale=0.425]{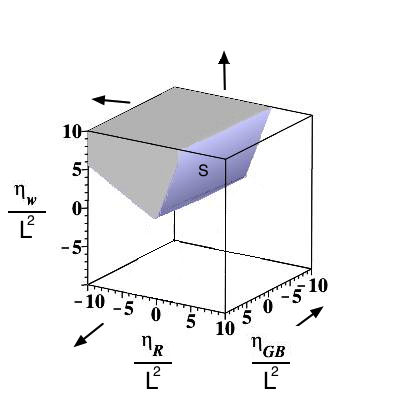}
		\caption{$d=3$, $z=2$, $\theta=2$.} \label{fig:etaGB_etaR_etaW_d3_z2_theta2}
	\end{subfigure}
\caption{Restrictions on $\eta_{R}$, $\eta_{GB}$ and $\eta_{W}$ from the NEC for $d=3$ and $z=2$, and $\theta = 1$ and $\theta = 2$, respectively. Note the change in the allowed region when going from $\theta = 1$ to $\theta = 2$.}{\label{fig:etaGB_etaR_etaW_d3_z2_theta1_2}}
\end{figure}

The conditions we need to satisfy are
 \begin{align}{\label{D4RGBW}}
 &-\eta_{GB} (d-1)(d-2)(d+1)(d-\theta)^{3}(d(z-1)-\theta) \notag \\
 &-\eta_{R}(d+1)\{d^{3}(z-1)-dz\theta (d-2) + (d-8)\theta^{2}\} \notag \\
 &\times[(d+1)\theta^{2} - 2(d+1)z\theta -2(d+1)d\theta + d^{2} (d+1) + 2d(d+z)z] \notag \\
 &+2\eta_{W}z(z-1)d(d-1)\{d(d-\theta)+2\theta\}[d(d-\theta) - d(z-2) + 2\theta)] \ge 0 \,,
 \end{align}
 \begin{align}{\label{D6RGBW}}
 &-\eta_{GB}(z-1)(d-1)(d-2)(d+1)(d-\theta)^{2} [d(z-\theta+d) + 2\theta] \notag \\
 &-\eta_{R}d(d+1)(z-1)\left\{d(z-\theta+d) + 2\theta\right\} \times \notag \\
 & \qquad\qquad\qquad\times [(d+1)\theta^{2} - 2(d+1)z\theta -2(d+1)d\theta + d^{2} (d+1) + 2d(d+z)z]  \notag \\
 &+2\eta_{W} z(z-1)(d-1)\{d(z-\theta+d)+2\theta\}[d^{2}(d+2) + (d+2)\theta-d^{2}(z+\theta)] \ge 0 \,,
 \end{align}
which are to be supplemented by (\ref{NEC1}) and (\ref{NEC2}). Once again, for $d=1$ and $d=2$, the Gauss-Bonnet term does not contribute, so we will begin our analysis at $d=3$.  In this case, we have three $\eta$s but only two conditions to satisfy, hence there is a wide range of possible values to choose from.  Nevertheless, we can still generate markedly different behaviour by change the value of the hyperscaling violation parameter, $\theta$.  Figure (\ref{fig:etaGB_etaR_etaW_d3_z2_theta1_2}) below provides an example for $d=3$, $z=2$ and $\theta = 1$ and $\theta = 2$, respectively.  In going from $\theta = 1$ to $\theta = 2$, a wide range of possible combinations of the $\eta$s is lost. 
 
\vspace{-12pt}
%======FIGURE======%
%======FIGURE======%
\vspace{-12pt}
 
\begin{figure}[H]
\centering
	\begin{subfigure}{0.49\linewidth} \centering
		\includegraphics[scale=0.425]{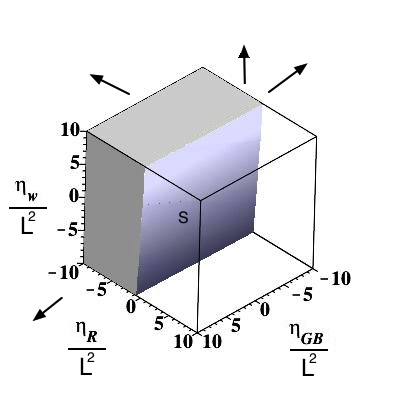}
		\caption{$d=3$, $z=4$, $\theta = 1$.} \label{fig:etaGB_etaR_etaW_d3_z4_theta1}
	\end{subfigure}
	\begin{subfigure}{0.49\linewidth} \centering
		\includegraphics[scale=0.425]{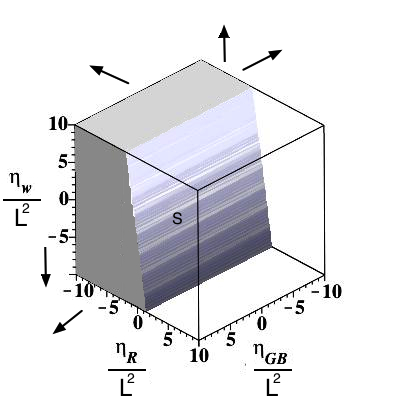}
		\caption{$d=3$, $z=6$, $\theta=2$.} \label{fig:etaGB_etaR_etaW_d3_z6_theta2}
	\end{subfigure}
\caption{Restrictions on $\eta_{R}$, $\eta_{GB}$ and $\eta_{W}$ from the NEC for $d=$3, $z=4$, $\theta = 1$ and $d=3$, $z=6$, $\theta=2$, respectively.}{\label{fig:etaGB_etaR_etaW_d3_z4_6}}
\end{figure}

The behaviour changes once again around $z=4$, this is depicted in 
Fig.\ref{fig:etaGB_etaR_etaW_d3_z4_6} for $d=3$.  For larger values of $z$, the allowed region is qualitatively similar to that of $z=4$, this is also depicted in Fig.\ref{fig:etaGB_etaR_etaW_d3_z4_6}.

\vspace{-12pt}
%======FIGURE======%
%======FIGURE======%
\vspace{-12pt}

\begin{figure}[H]
\centering
	\begin{subfigure}{0.3\linewidth} \centering
		\includegraphics[scale=0.425]{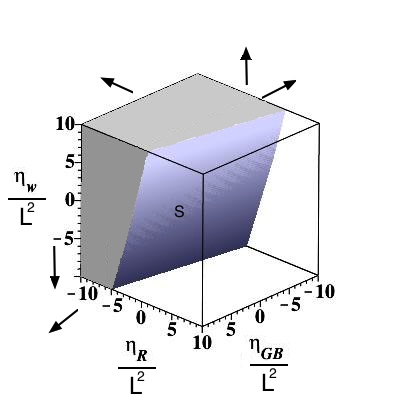}
		\caption{$d=5$, $z=2$, $\theta = 1$.} \label{fig:etaGB_etaR_etaW_d5_z2_theta1}
	\end{subfigure}
	\hskip0.03\linewidth
	\begin{subfigure}{0.3\linewidth} \centering
		\includegraphics[scale=0.425]{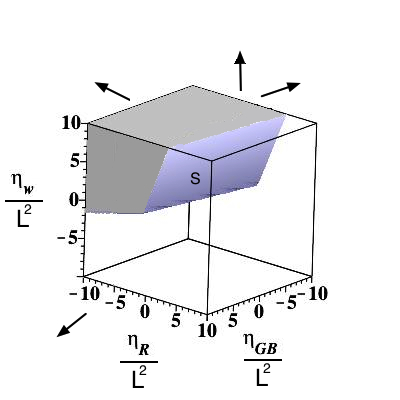}
		\caption{$d=5$, $z=2$, $\theta=3$.} \label{fig:etaGB_etaR_etaW_d5_z2_theta3}
	\end{subfigure}
	\hskip0.03\linewidth
	\begin{subfigure}{0.3\linewidth} \centering
		\includegraphics[scale=0.425]{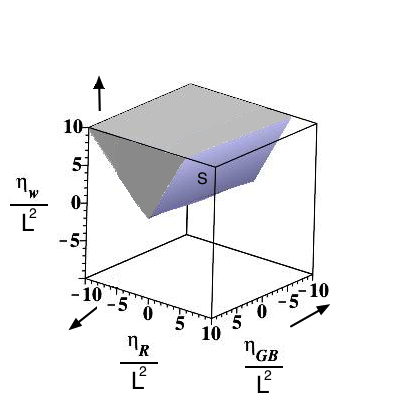}
		\caption{$d=5$, $z=2$, $\theta=4$.} \label{fig:etaGB_etaR_etaW_d5_z2_theta4}
	\end{subfigure}
\caption{Restrictions on $\eta_{R}$, $\eta_{GB}$ and $\eta_{W}$ from the NEC for $d=5$ and $z=2$, and $\theta = 1$, $\theta = 3$, and $\theta=4$, respectively. Note the change in the allowed region when going from $\theta = 1$ to $\theta = 3$ and $\theta=4$.}{\label{fig:etaGB_etaR_etaW_d5_z2_theta123}}
\end{figure}

\begin{figure}[H]
\centering
	\begin{subfigure}{0.49\linewidth} \centering
		\includegraphics[scale=0.425]{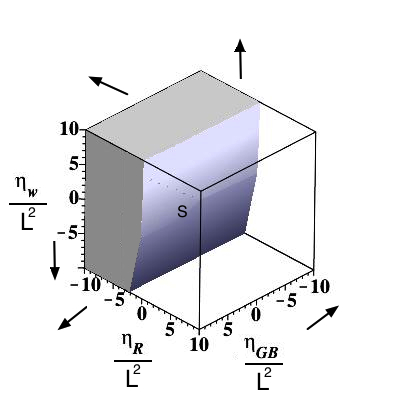}
		\caption{$d=5$, $z=4$, $\theta = 4$.} \label{fig:etaGB_etaR_etaW_d5_z4_theta4}
	\end{subfigure}
	\begin{subfigure}{0.49\linewidth} \centering
		\includegraphics[scale=0.425]{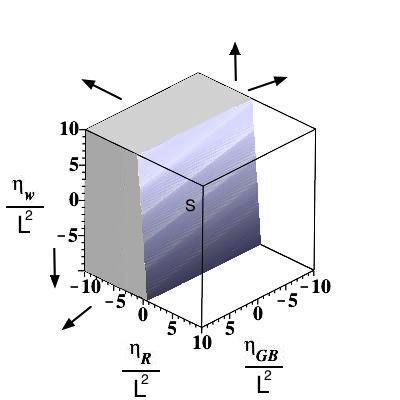}
		\caption{$d=5$, $z=6$, $\theta=3$.} \label{fig:etaGB_etaR_etaW_d5_z6_theta3}
	\end{subfigure}
	\caption{Restrictions on $\eta_{R}$, $\eta_{GB}$ and $\eta_{W}$ from the NEC for $d=$5, $z=4$, $\theta = 4$ and $d=5$, $z=6$, $\theta=3$, respectively.}{\label{fig:etaGB_etaR_etaW_d5_z2_6}}
\end{figure}

Again, qualitatively similiar results are obtained for higher dimensions.  Below $z=4$, the allowed region of $\eta$s can be quite different, depending on the value of $\theta$.  Above, $z\ge4$, the shape of the allowed region is not as sensitive to changes of $\theta$.  Figures (\ref{fig:etaGB_etaR_etaW_d5_z2_theta123}) and (\ref{fig:etaGB_etaR_etaW_d5_z2_6}) provide examples for $d=5$.  

\vspace{-12pt}
%======FIGURE======%
%======FIGURE======%
\vspace{-12pt}

\vspace{-12pt}
%======FIGURE======%
%======FIGURE======%
\vspace{-12pt}
 
We will summarize the general features of all these NEC plots and their physical implications in Section \ref{sec:discussion}. 
 
%========+=========+=========+=========+=========+=========+=========+=========+
\section{Crossover solutions}{\label{sec:crossover}}

In this section we are interested in constructing completions to the $D=d+2$ HSV geometry in the deep interior (IR) and asymptotic region (UV). We will specialize to Einstein-Weyl gravity because the Weyl tensor vanishes in $AdS$, which means that the curvature squared terms will not source the dilaton regardless of the form of its coupling. This ensures that we have a chance at an $AdS$ UV completion to the geometry. We also seek an IR completion to $AdS_{2} \times \mathbb{R}^{d}$ in the IR so that any gravitational (tidal force)  singularities in the deep interior of the HSV spacetime are resolved by the crossover to the singularity-free $AdS_{2} \times \mathbb{R}^{d}$. We also hope to tame potential curvature invariant blowups in the asymptotic region, in this case by the $AdS_D$. This type of analysis was performed for the Lifshitz case in \cite{lifcorrect}, which provided one of the motivations for the analysis of this section, but it uses a different set of theory functions than ours.  

The HSV solution that was contructed in Section \ref{sec:r2hsv} contains a running dilaton.  If we want to be able to complete the spacetime into an $AdS_{2} \times \mathbb{R}^{d}$ in the IR, we need a mechanism by which the dilaton is stabilized to some constant value, $\phi_{0}$.  This is where the curvature squared corrections to the action come into play. 
An immediate question that arises is whether we need to add any further terms to our effective action in order to support crossovers. After allowing corrections to the Lagrangian of the form $k(\phi)C_{\mu\nu\lambda\sigma}C^{\mu\nu\lambda\sigma}$, we find that it is possible to have HSV as a solution of the equations as well as $AdS_D$ and $AdS_2\times \mathbb{R}^d$ without needing any more theory functions than $f(\phi)$ and $V(\phi)$. The details are messy and best left suppressed.

It is most sensible for us to seek crossovers to HSV starting from the $AdS_D$ and $AdS_2\times\mathbb{R}^d$ ends, aiming to pick up the HSV solution as we evolve in radius in between. This is because classification of relevant/irrelevant perturbations is much better understood in $AdS$.
 
In order to find crossovers, it helps to pick a convenient gauge for the metric (\ref{hsvintro}).  Under a coordinate transformation ${\mathcal{R}} = \rho^{-1/z}$ and a rescaling of $L$ and the other coordinates, the line element takes the form
\begin{equation}{\label{last_hsv}}
ds^{2} = L^{2} \rho^{-2(dz-\theta)/dz} \left[ -dt^{2} + d \rho^{2}  + \rho^{2(z-1)/z} d{\vec{x}}_{d}^2 \right] \,.
\end{equation}
In the new $\rho$ coordinate, the deep interior (IR) corresponds to $\rho \rightarrow \infty$ and the UV corresponds to $\rho \rightarrow 0$.  Since we wish to construct solutions which flow from $AdS_{2} \times \mathbb{R}^{d}$ in the IR to and intermediate HSV regime and finally to $AdS_D$, it is convenient to parametrize the metric as \cite{lifcorrect}:
\begin{equation}{\label{metric_shooting}}
ds^{2} = a^{2}_{1}(\rho) \left[ 
- dt^{2} + d\rho^{2}+ a^{2}_{2}(\rho)d{\vec{x}}_d^2 \right] \,.
\end{equation}
In this parametrization, $AdS_{2} \times \mathbb{R}^{d}$ corresponds to $a_{1}(\rho) = {{L_\flat}}/\rho$ and $a_{2}(\rho) =  \rho$, whereas $AdS_{D}$ corresponds to $a_1(\rho)=  L_\sharp/\rho$ and $a_{2}(\rho)=\, $const.   

In what follows we denote radial coordinate derivatives by $' = \partial_{\rho}$. We will also abbreviate dilaton field derivatives of theory functions as ${\dot{\ }}\equiv \partial_\phi$. This is not a time derivative; instead, it is a {\em field} derivative. We trust the reader not to get confused by this.

It is easiest to begin our study of the field equations with the gauge field.
The Maxwell field equation is
\begin{equation}{\label{shooting_maxwell}}
\nabla_{\mu} \left[ f(\phi) F^{\mu \nu} \right] = 0 \,.
\end{equation}
In our metric ansatz, this equation becomes
\begin{equation}
\left[ a_1^{d+2} a_2^{d} f(\phi) {\cal{E}} \right]' = 0 \,,
\end{equation}
where ${\cal{E}}\equiv F^{\rho t}(\rho)$ is a solution function of $\rho$. Clearly, this equation has a first integral, 
\begin{equation}
{\cal{E}}(\rho) = {\frac{Q}{a_1^{d+2} a_2^{d} f(\phi)}} \,,
\end{equation}
where $Q$ is an integration constant. Note that there is a theory function involved here: $f(\phi)$. Note also that although $Q$ is a constant, it will be a {\em different} constant for our three different solutions involved in the crossover: $AdS_D$, $AdS_2\times \mathbb{R}^d$, and HSV. This is related to the fact that the electric field must have perturbations if the dilaton and metric do, in accordance with the terms in the original effective action. It is the total electric field which obeys the Maxwell equation: the background, about which one is expanding, plus the perturbation. (Note that if we had done magnetic perturbations staying in a radial ansatz, the perturbed magnetic field would be unconstrained by the Maxwell equation.) The last ingredient we need is $\partial_\rho f(\phi)$. This is directly available because we are working within an ansatz with {only} radial coordinate dependence, so 
\begin{equation}
[f(\phi(\rho))]'= \frac{df}{d\phi} {\frac{d\phi}{d\rho}} \,,
\end{equation}
or, more succinctly, $f'={\dot{f}}\phi'$.
This gives the full Maxwell equation as
\begin{equation}\label{Maxwell_shooting}
\left[ (d+2){\frac{a_1''}{a_1}} + d{\frac{a_2''}{a_2}} \right] {\cal{E}}
+ {\frac{{\dot{f}}(\phi)}{f(\phi)}}\phi' {\mathcal{E}} + {\mathcal{E}}' =0 \,.
\end{equation}

The dilaton field equation is
\begin{equation} 
\Box \phi - {\dot{f}}(\phi) F_{\mu \nu} F^{\mu \nu} - {\dot{V}}(\phi) = 0 \,.
\end{equation}
When evaluated on the metric ansatz (\ref{metric_shooting}), this gives the full  dilaton equation as
\begin{equation}{\label{dilaton_shooting}}
\frac{1}{a_{1}^{2}} \phi'' + \frac{d}{a_{1}^{2}} \left ( \frac{a_{1}'}{a_{1}} + \frac{a_{2}'}{a_{2}} \right ) \phi' - {\dot{V}}(\phi) 
+ 2 {\dot{f}}(\phi) a_1^4 {\mathcal{E}}^2  = 0 \,.
\end{equation}
These coupled equations (\ref{Maxwell_shooting}) and (\ref{dilaton_shooting}) for $\cal{E}$ and $\phi$ show us that it is impossible to turn on dilaton perturbations without also exciting electric perturbations. 

The next step is to write down the energy-momentum tensor for the right hand side of the tilded gravitational equations of motion. The components of $T^\mu_{\ \nu}$ are
\begin{align}{\label{EnMon_shooting_tt}}
T^t_{\ t} = & -{\frac{1}{2}} V(\phi) - {\frac{1}{4a_1^2}}(\phi')^2 
- {\frac{1}{2}}f(\phi)a_1^4{\mathcal{E}}^2 \,,
\end{align}
\begin{equation}{\label{EnMon_shooting_rr}}
T^\rho_{\ \rho} = -{\frac{1}{2}} V(\phi) + {\frac{1}{4a_1^2}}(\phi')^2 
- {\frac{1}{2}}f(\phi)a_1^4{\mathcal{E}}^2 \,,
\end{equation}
\begin{align}{\label{EnMon_shooting_ii}}
T^x_{\ x} & = -{\frac{1}{2}} V(\phi) - {\frac{1}{4a_1^2}}(\phi')^2 
+ {\frac{1}{2}}f(\phi)a_1^4{\mathcal{E}}^2 \,.
\end{align}
For the $\widetilde{G}_{\mu\nu}$s in the Weyl corrected gravity equations $\widetilde{G}_{\mu\nu}=T_{\mu\nu}$, we write the second and fourth order pieces as  
\begin{align}{\label{ein_shooting_tt}}
\widetilde{G}_{t t} \equiv  &
+ c_1^t \left( \frac{a_{1}'}{a_{1}} \right )^{2} 
+ c_2^t \left(\frac{a_{2}'}{a_{2}} \right )^{2} 
+ c_3^t \frac{a_{1}''}{a_{1}}
+ c_4^t \frac{a_{2}''}{a_{2}} 
+ c_5^t \frac{a_{1}'}{a_1} \frac{a_{2}'}{a_{2}}
\notag \\ &
+\frac{4 \eta_W}{3 a_{1}^{2}} \left \{ 
+ c_6^t \left (\frac{a_{2}'}{a_{2}} \right )^{4} 
+ c_7^t \frac{a_{2}''''}{a_{2}} 
+ c_8^t \frac{a_2''}{a_2}\left({\frac{a_2'}{a_2}}\right)^2
+ c_9^t \frac{a_{2}'''}{a_{2}} \frac{a_{2}'}{a_2}  
+ c_{10}^t \left( \frac{a_{2}''}{a_{2}} \right)^2 
\right. \notag\\ & \left. \qquad\qquad
+ c_{11}^t {\frac{a_2'''}{a_2}}\frac{a_{1}'}{a_1}
+ c_{12}^t \frac{a_{1}''}{a_{1}}\left(\frac{a_{2}'}{a_{2}} \right )^{2} 
+ c_{13}^t \frac{a_{1}'}{a_{1}}\left(\frac{a_{2}'}{a_{2}} \right )^{3}
+ c_{14}^t \frac{a_{1}''}{a_{1}}\frac{a_{2}''}{a_{2}}
\right. \notag\\ & \left. \qquad\qquad
+ c_{15}^t \frac{a_2''}{a_2}\frac{a_{1}'}{a_{1}}\frac{a_{2}'}{a_{2}}
+ c_{16}^t \left(\frac{a_{1}'}{a_{1}}\right )^{2}
            \left(\frac{a_{2}'}{a_{2}}\right)^{2}
+ c_{17}^t \left(\frac{a_{1}'}{a_{1}}\right )^{2}\frac{a_{2}''}{a_{2}}
\right \}  \,,
\end{align}
and similarly for the $\widetilde{G}_{\rho\rho}$, and for $a_2^{-2} \, \widetilde{G}_{xx}$ for each coordinate in the ${\vec{x}}$. Note that at fourth order, in principle there might have been three other types terms of the form $(a_1''''/a_1)$, $(a_1'''/a_1)(a_1'/a_1)$, $(a_1'''/a_1) (a_2'/a_2)$, but these are absent in conformal gauge. 

The 17 constants $\{c^t_{I}\}$ in these expressions are all functions of $d$, and similarly with the  $\{c^\rho_{I}\}$ and $\{c^x_{I}\}$. We now list them.
The coefficients of the second order bits for $\widetilde{G}_{tt}$ are
\begin{align}
 c_1^t &= -{\frac{1}{2}}d(d-3) \,, &
 c_2^t &= -{\frac{1}{2}}d(d-1)  \,, &
 c_3^t &= -d \,,& \notag\\ 
 c_4^t &= -d \,,& 
 c_5^t &= -d^2 \,;&
 \quad & \quad 
\end{align}
while the fourth order bits are
\begin{align}
 c_6^t    &= +{\frac{3(d-1)[2d^2-8d+7]}{2(d+1)}}\,,& 
 c_7^t    &= -{\frac{3(d-1)}{(d+1)}}\,,& & \notag\\
 c_8^t    &= -{\frac{3(d-1)[d^2-7d+8]}{(d+1)}}\,,&
 c_9^t    &= -{\frac{3(d-1)(2d-3)}{(d+1)}}\,,&\notag\\
 c_{10}^t &= -{\frac{3(d-1)(2d-5)}{2(d+1)}}\,,&
 c_{11}^t &= -{\frac{6(d-1)(d-2)}{(d+1)}}\,,& \notag\\
 c_{12}^t &= -{\frac{1}{2}}c_{11}^t\,,&
 c_{13}^t &= +{\frac{3(d-1)(d-2)(2d-3)}{(d+1)}}\,,&\notag\\
 c_{14}^t &= +{\frac{1}{2}}c_{11}\,,& 
 c_{15}^t &= -{\frac{3(d-1)(d-2)(2d-5)}{(d+1)}}\,,&\notag\\
 c_{16}^t &= +{\frac{3(d-1)(d-2)(d-3)}{(d+1)}}\,,&
 c_{17}^t &= -c_{16}^t\,.&
\end{align}
The coefficients of the second order bits for $\widetilde{G}_{\rho\rho}$ are
\begin{align}
 c_1^\rho &= +{\frac{d(d+1)}{2}}\,,&
 c_2^\rho &= +{\frac{d(d-1)}{2}}\,,&
 c_3^\rho &= 0\,,& \notag\\ 
 c_4^\rho &= 0\,,& 
 c_5^\rho &= +d^2\,;&
 \quad & \quad 
\end{align}
while the fourth order bits are
\begin{align}
 c_6^\rho    &= +{\frac{3(d-1)(2d-3)}{2(d+1)}}\,,& 
 c_7^\rho    &= 0\,,& 
 c_8^\rho    &= -{\frac{3(d-1)(d-2)}{(d+1)}}\,,&  
 c_9^\rho    &= -{\frac{3(d-1)}{(d+1)}}\,,& \notag\\
 c_{10}^\rho &= -{\frac{1}{2}}c_9^\rho\,,&
 c_{11}^\rho &= 0\,,& 
 c_{12}^\rho &= 0\,,&
 c_{13}^\rho &= -c_8^\rho\,,& \notag\\
 c_{14}^\rho &= 0\,,& 
 c_{15}^\rho &= -c_8^\rho\,,& 
 c_{16}^\rho &= 0\,,& 
 c_{17}^\rho &= 0\,.&
\end{align}
The coefficients of the second order bits for  each of the $(a_2^{-2})\widetilde{G}_{xx}$ are
\begin{align}
 c_1^x &= +{\frac{1}{2}}d(d-3) \,,&
 c_2^x &= +{\frac{1}{2}}(d-1)(d-2) \,,&
 c_3^x &= +d \,,& \notag\\ 
 c_4^x &= +(d-1)\,,& 
 c_5^x &= +d(d-1)\,;&
 \quad & \quad 
\end{align}
while the fourth order bits are
\begin{align}
 c_6^x    &= +{\frac{3(d-1)(d-4)(2d-3)}{2d(d+1)}}\,,& 
 c_7^x    &= -{\frac{3(d-1)}{d(d+1)}}\,,& \notag\\
 c_8^x    &= -{\frac{3(d-1)[d^2-9d+12]}{d(d+1)}} \,,& 
 c_9^x    &= -6{\frac{(d-1)(d-2)}{d(d+1)}} \,,& \notag\\
 c_{10}^x &= +{\frac{3}{4}}c_9^x\,,&
 c_{11}^x &= +c_9^x\,,& \notag\\
 c_{12}^x &= -{\frac{1}{2}}c_9^x\,,&
 c_{13}^x &= +{\frac{6(d-1)(d-2)^2}{d(d+1)}}\,,& \notag\\
 c_{14}^x &= +{\frac{1}{2}}c_9^x\,,&  
 c_{15}^x &= -2c_{16}^x\,,& \notag\\
 c_{16}^x &= +{\frac{3(d-1)(d-2)(d-3)}{d(d+1)}}\,,&
 c_{17}^x &= -c_{16}^x\,.&
\end{align}

Now that we have the full equations of motion, we can set about demanding that the three different spacetimes which we want to participate in the crossover solve the equations of motion: (1) $AdS_D$, (2) $AdS_2\times \mathbb{R}^d$, and (3) HSV. We next consider these in turn.

First, we need to verify that $AdS_D$ is a solution of our Einstein-Weyl-Maxwell-dilaton action. $AdS_D$ has $a_2=1$ and $a_1={{L}_\sharp}/\rho$. Let the constant dilaton in the $AdS_D$ region be $\phi_\sharp$, along with $V_\sharp\equiv V(\phi_\sharp)$, $f_\sharp\equiv f(\phi_\sharp)$, and similarly for higher derivatives of theory functions. Then the equations of motion for the metric coefficients evaluated on $AdS_D$ (in $D$ dimensions) yield two conditions,
\begin{equation}
Q^2_\sharp=0 \,, \qquad V_\sharp=-{\frac{d(d+1)}{{L_\sharp}^2}} \,.
\end{equation}

Next, let us see whether $AdS_2\times \mathbb{R}^d$ is also a solution. This has $a_{1}(\rho) = L/\rho$ and $a_{2}(\rho) = \rho$, and $\phi(\rho) = \phi_{\flat} =\ $const. Adding and subtracting two Einstein equations yields 
\begin{equation}
Q^{2}_\flat = L_\flat^{2(d-1)} f(\phi_{\flat}) 
\left[1 - \frac{4(d-1)}{(d+1)L_\flat^{2}} \eta_W \right]
\,, \qquad V_\flat = -\frac{1}{L_\flat^{2}} \,.
\end{equation}
As expected, $V(\phi_{\flat})$ sets the scale for the $AdS_{2} \times \mathbb{R}^{d}$ solution. Notice that the presence of the Weyl squared term changes the effective charge compared to Einstein gravity, reducing or increasing it depending on the sign of $\eta_W$. 

Lastly, we can ask whether HSV is a solution. This was already ensured by design in Section \ref{sec:r2hsv}.

In linearized perturbation theory, the Maxwell field equation guarantees that the electric field will be perturbed along with the dilaton. We expand
\begin{equation}
{\mathcal{E}}(\rho) = \bar{\mathcal{E}}(\rho) + E(\rho) \,,
\end{equation}
where $\bar{\mathcal{E}}(\rho)$ is the background electric field and $E(\rho)$ the perturbation. We also split the dilaton into a background piece $\bar{\phi}(\rho)$ (which will be just a constant for $AdS_D$ and $AdS_2\times \mathbb{R}^d$) and a perturbation $\Phi$,
\begin{equation}
\phi(\rho) = \bar{\phi}(\rho) + \Phi(\rho) \,.
\end{equation}
%
%The metric coefficients are split into background pieces $\bar{a}_1, \, \bar{a}_2$ and perturbations $ A(\rho), \, B(\rho)$, respectively,
%
%\begin{equation}
%a_1 = \bar{a}_1 + L\,A(\rho), \, \, \, a_2 = \bar{a}_2 + B(\rho) \, .
%\end{equation}
%
Using the form of our metric ansatz, the linearized electric equation becomes
\begin{equation}
\left[ 
(d+2){\frac{{\bar{a}}_1'}{\bar{a}_1}} + d{\frac{\bar{a}_2'}{\bar{a}_2}} \right]
\left\{ f_0 E + {\dot{f}}_0 \bar{\mathcal{E}} \Phi \right\}
+{\dot{f}}_0\bar{\mathcal{E}} \Phi' + f_0 E' = 0 \,,
\end{equation}
where $f_0$ denotes the theory function $f(\phi)$ evaluated at its background value and $\bar{a}_1, \, \bar{a}_2$ are the background metric coefficients. The linearized dilaton perturbation equation becomes 
\begin{equation}
{\frac{1}{{\bar{a}}_1^2}}\phi'' 
+  {\frac{d}{\bar{a}_1^2}}  \left[
{\frac{{\bar{a}}_1'}{{\bar{a}}_1}} + \frac{\bar{a}_2'}{\bar{a}_2} \right] \phi'
 - {\dot{V}}_0 - {\ddot{V}}_0 \Phi
+ 2 {\bar{a}}_1^4 \left[ \bar{\mathcal{E}}^2\left( {\dot{f}}_0 
+ {\ddot{f}}_0 \Phi \right) + 2{\dot{f}}_0 \bar{\mathcal{E}} E \right] =0 \,.
\end{equation}

We can also write the linearized energy-momentum tensor for use in finding crossovers. Since at linear order the $(\Phi')^2$ pieces drop out,
\begin{equation}
\left. T^t_{\ t} \right|_{\rm lin} 
 = -{\frac{1}{2}}{{V}}_0 - {\frac{1}{2}}{\dot{V}}_0 \Phi 
- {\frac{Q_0}{2{{\bar{a}}_1^{d-2}\bar{a}_2^{d}}}} \left\{ 
{\frac{Q_0}{{\bar{a}}_1^{d+2} {\bar{a}}_2^{d} f_0}} \left(
1 + {\frac{{\dot{f}}_0}{f_0}} \Phi + {\frac{4AL}{{\bar{a}}_1}} \right)
+2E \right\}   \,,
\end{equation}
and
\begin{equation}
\left. T^t_{\ t} \right|_{\rm lin} 
 = \left. T^\rho_{\ \rho}\right|_{\rm lin} \,,
\end{equation}
while 
\begin{equation}
\left. T^x_{\ x} \right|_{\rm lin}  
= -{\frac{1}{2}}{{V}}_0 - {\frac{1}{2}}{\dot{V}}_0 \Phi 
+ {\frac{Q_0}{2{{\bar{a}}_1^{d-2}\bar{a}_2^{d}}}} \left\{ 
{\frac{Q_0}{{\bar{a}}_1^{d+2} {\bar{a}}_2^{d} f_0}} \left(
1 + {\frac{{\dot{f}}_0}{f_0}} \Phi + {\frac{4AL}{{\bar{a}}_1}} \right)
+2E \right\}   \,,
\end{equation}
where $A = A(\rho)$ is a metric perturbation: $a_1 = \bar{a}_1 + L \, A(\rho)$. The form of the metric perturbation equations depends on the background around which we expand. In particular, the analysis differs for $AdS_D$ and $AdS_2\times \mathbb{R}^d$, so we split the discussion at this point.

%========+=========+=========+=========+=========+=========+=========+=========+
\subsection{UV crossover}

The perturbed $AdS_D$ metric is in conformal gauge 
\begin{equation}
ds^2_{AdS_D} = \left\{ {\frac{L_\sharp}{\rho}} + L_\sharp A(\rho) \right\}^2 \left[ -dt^2 + d\rho^2 + \left\{ 1 + B(\rho) \right\}^2 d{\vec{x}}^2_d \right] \,.
\end{equation}
We now expand the curvature squared equations of motion at linear order in perturbations $\{A(\rho),B(\rho),\Phi(\rho),E(\rho)\}$. We obtain for the $tt$ gravity field equation
\begin{align}\label{ads4lintt}
0 &=  
- 4(d-1) {\frac{\eta_W}{L_\sharp^2}} (\rho^4 B'''')
+ 8(d-1)(d-2) {\frac{\eta_W}{L_\sharp^2}} (\rho^3 B''')
\notag\\ & \quad
+ \left[ -d(d+1) - 4(d-1)^2(d-2) {\frac{\eta_W}{L_\sharp^2}} \right] 
  (\rho^2 B'')
-d(d+1) (\rho^3 A'') 
\notag\\ & \quad
+d^2(d+1)(\rho B')
+d(d+1)(d-3) (\rho^2 A') 
+2d^2(d+1)(\rho A) \,,
\end{align}
while for $\rho\rho$ we obtain
\begin{equation}\label{ads4linpp}
0= d(\rho B') + (d+1)(\rho^2 A') + 2(d+1)(\rho A)
\,,
\end{equation}
while for (each) $xx$ we get
\begin{align}\label{ads4linxx}
0 &= 
+{\frac{4(d-1)}{d}} {\frac{\eta_W}{L_\sharp^2}} (\rho^4 B'''')
-{\frac{8(d-1)(d-2)}{d}} {\frac{\eta_W}{L_\sharp^2}} (\rho^3 B''')
\notag\\ &
\left[ -(d-1)(d+1)  + {\frac{4(d-1)^2(d-2)}{d}}{\frac{\eta_W}{L_\sharp^2}} 
  \right] (\rho^2 B'')
-d(d+1)  (\rho^3 A'')
\notag\\ &
+d(d-1)(d+1) (\rho B')
+d(d+1)(d-3)  (\rho^2 A') 
+2d^2(d+1) (\rho A) \,.
\end{align}
The linearized Maxwell equation is simple by dint of gauge symmetry,
\begin{equation}\label{ads4linE}
E' - {\frac{(d+2)}{\rho}}E = 0 \,.
\end{equation}
The linearized dilaton equation of motion (\ref{dilaton_shooting}) in the perturbed $AdS_D$ background is also quite simple,
\begin{equation}\label{ads4linP}
\Phi'' - {\frac{d}{\rho}}\Phi' 
- {\frac{L_\sharp^2}{\rho^2}} {\ddot{V}}_\sharp \,\Phi =0   \,.
\end{equation}

The form of our perturbation equations (\ref{ads4lintt}-\ref{ads4linP}) permits {\rm power law} solutions of the form
\begin{equation}
\Phi  = \breve{\Phi} \rho^{\nu_\Phi} \,, \qquad
A =  \breve{A} \rho^{\nu_A} \,,\qquad
B =  \breve{B} \rho^{\nu_B} \,,\qquad
E =  \breve{E} \rho^{\nu_E} \,,
\end{equation}
where $\{\breve{A},\breve{B},\breve{\Phi},\breve{E} \}$ are simple constants, as long as two conditions are satisfied relating the metric indices $\nu_A, \nu_B$ to $\nu_\Phi$,
\begin{align}\label{ads4nuAB}
\nu_A & =\nu_B -1  \,, \\
\nu_\Phi & = \nu_B \,.
\end{align}

The electric field index $\nu_E$ is determined by the first order Maxwell equation to be
\begin{equation}\label{ads4nuE}
\nu_E = d+2\,.
\end{equation}
In other words, there is only one perturbation of the electric field and it is relevant (growing in the IR). This is in accord with our intuition that $Q$ should evolve from $Q_\sharp=0$ in the asymptotic $AdS_D$ region, increasing in the interior \`a la HSV, and eventually levelling out to $Q_\flat$ of the interior $AdS_2\times \mathbb{R}^d$. 

The second order dilaton equation of motion determines the two allowed values of
the index $\nu_\Phi$, 
\begin{equation}\label{ads4nuP}
\nu_\Phi^\sharp = {\frac{(d+1)}{2}}\pm 
\sqrt{{\frac{(d+1)^2}{4}+L_\sharp^2{\ddot{V}}_\sharp}} \,.
\end{equation}
Here, we see two perturbations, one relevant and one irrelevant. This simple equation (\ref{ads4nuP}) is the familiar one from $AdS$, with ${\ddot{V}}_\sharp$ playing the role of $m^2$, as we would expect by consistency. 

Morally, we must inspect the form of our theory function $V(\phi)$ to check that  it gives rise to {\em real} $\nu_\Phi$: we certainly do not want oscillatory solutions indicating a (linearized) instability. We require the term under the square bracket to be non-negative,
\begin{equation}
\diamondsuit_\sharp \equiv {\frac{(d+1)^2}{4}}+L_\sharp^2{\ddot{V}}_\sharp \geq 0 \,.
\end{equation}
Let us check what kind of theory parameters can support real dilaton perturbations. We assume for simplicity that $V(\phi)$ and $f(\phi)$ are monotonic; if not, our story becomes more involved. 
Because we work within a purely radial ansatz and assume monotonicity of $V(\phi),f(\phi)$, we can find $\ddot{V}$ from our implicit expressions in Section \ref{sec:r2hsv},
\begin{equation}
{\ddot{V}} = \left( {\frac{d\phi}{dr}} \right)^{-3} \left\{ {\frac{d^2V}{dr^2}}{\frac{d\phi}{dr}} - {\frac{dV}{dr}}{\frac{d^2\phi}{dr^2}} \right\}  \,.
\end{equation}
Previously, we found that our theory could only support HSV solutions for particular classes of functions $f(\phi),V(\phi)$ whose  coefficients are specified by six constants $\{ D_1,\ldots, D_6 \}$. This eventuates because from the structure of the effective action: there cannot be derivatives of theory functions higher than second order, so to specify $f_0,{\dot{f}}_0, {\ddot{f}}_0, V_0, {\dot{V}}_0, {\ddot{V}}_0$ we will need exactly six variables. Let us collect the relevant facts here about the $\{D_i\}$ that follow from the NEC, causality, and physical $d_{\rm eff}$. The even constants $D_2,D_4,D_6$ are the only ones that depend on $\eta_W$, and they all depend on it linearly. By virtue of the NEC, $D_3,D_4,D_5,D_6$ are all positive. Also, in the physical ranges of $z,\theta$, $D_1$ is positive. The most interesting constant is $D_2$. It can be positive, negative, or zero, and its behaviour depends strongly on whether or not $d=2$ or $d>2$. For $d=2$, $D_2=0$ if $z=4$ for any $\theta$. For $d>2$, $D_2=0$ if $\theta=-d(dz-z-2d)/(d-\theta)$ so for positive $\theta$ (the physical range) we need $z<2d/(d-1)$. With all that noted, we now have all the ingredients necessary to rule out tachyons. 

Examining $L_\sharp^2 {\ddot{V}}_\sharp$ in the asymptotic region, we can notice something immediately. It is some constants of order one multiplied by one factor of $\eta_W/L_\sharp^2$, a parameter which must be small. Therefore, in our regime of theory parameters, the $\Phi$ perturbation is not in danger of becoming tachyonic because $\eta_W/L_\sharp^2$ is tiny. 
We must also inspect all positive powers of $r$ in $L_\sharp^2\ddot{V}_\sharp$ in the $AdS_D$ regime, to ensure that their coefficient(s) go to zero in the UV region. Having a finite $L_\sharp^2\ddot{V}_\sharp$ requires
\begin{equation}
D_2=0 \,.
\end{equation}
This requires either 
\begin{equation}
d=2: \quad z=4 \,, {\rm any}\ \theta \in [0,d) \,,
\end{equation}
or 
\begin{equation}
d>2: \quad z<{\frac{2d}{d-1}} \quad {\rm for}\ \ 
\theta={\frac{d(2d+z-dz)}{d-2}} >0 \,.
\end{equation}
This provides serious restrictions on the solution parameters in the physically interesting range.  
Thirdly, inspecting the constant term in $L_\sharp^2 {\ddot{V}}_\sharp$, which is now the dominant term at $r\rightarrow\infty$, we find that its coefficient is $-(\beta-1)^2[D_1D_4+ 8D_2D_3]$, which is proportional to $\eta_W$ (and terms of order one) and can be positive, negative, or zero depending on solution parameters. In one special case, $d=2$, this coefficient also turns out to be zero when $z=4$, making $\nu_\Phi$ extremely simple. 

Let us now outline the numerical shooting problem. In principle, with the equations of motion being fourth order in $B$, second order in $A,\Phi$, and first order in $E$, we would need to specify a set of nine items $\{E,\phi,\phi',A,A',B,B',B'',B'''\}$ to solve an initial value problem.  For our perturbation problem, we know the initial conditions for the perturbations (they are all zero), but not their derivatives, leaving five to shoot on numerically.
Now, notice that the sum of the $tt$ (\ref{ads4lintt}) and $xx$ (\ref{ads4linxx}) linearized gravity field equations produces an expression for $[(\rho^4 B'''')-2(d-2)(\rho^3B''')]$ in terms of $B''$ and $B'$ only -- all the terms involving $A'', A', A$ cancel out (and there were no $B$ terms to begin with). Substituting this back into the $tt$ gravity equation (\ref{ads4lintt}) reduces the order of the linearized differential equation from four to two. Looking back to the $\rho\rho$ linearized constraint equation (\ref{ads4linpp}), we can see that it is first order in $A$ and $B$,  which fits perfectly. The full nonlinear equations are still fourth order, but the dimensionality of the shooting problem is reduced by two: we need to shoot on only {\em three} derivatives $A',B',\Phi'$, just like in Einstein gravity. 
Note that this simplification will not persist for $AdS_2\times \mathbb{R}^d$; it is specific to the properties of $AdS_D$. Finally, to recognize the HSV metric while shooting, we would plot the logs of the metric coefficients and pick off $z,\theta$.

%========+=========+=========+=========+=========+=========+=========+=========+
\subsection{IR crossover}

This time the perturbed metric that is appropriate to $AdS_2\times \mathbb{R}^d$,
\begin{equation}
ds^2_{AdS_2\times\mathbb{R}^d} = \left\{ {\frac{L_\flat}{\rho}} + L_\flat A(\rho) \right\}^2 \left[ -dt^2 + d\rho^2 + \left\{ \rho + B(\rho) \right\}^2 \left(dx^2+dy^2 \right) \right] \,.
\end{equation}
Similarly to the $AdS_D$ case, we expand the curvature squared equations of motion at linear order in perturbations $\{A(\rho),B(\rho),\Phi(\rho),E(\rho)\}$. Note that these functions are not the same as for $AdS_D$ and the linearized perturbation equations will obviously differ. We obtain for the $tt$ equation
\begin{align}\label{ads2lintt}
& 0= + {\frac{4(d-1)}{(d+1)}}{\frac{\eta_W}{L_\flat^2}}
\left\{ (\rho^4 B'''') + (\rho^3 B''') \right\} 
+ \left[ d-{\frac{4d(d-1)}{(d+1)}} {\frac{\eta_W}{L_\flat^2}} \right] 
  \left\{ (\rho^2 B'')-(\rho B')\right\}
\notag\\ & \quad
+ \left[ d-{\frac{3d(d-1)(d-2)}{d(d+1)}}{\frac{\eta_W}{L_\flat^2}} \right] 
  \left\{ (\rho^4 A'') +3(\rho^3 A') \right\}  
\notag\\ & \quad
+ \left[ 1 - {\frac{4(d-1)}{(d+1)}} {\frac{\eta_W}{L_\flat^2}}\right] 
  \left\{ dB + (d+2) A\right\} 
+ f_\flat L_\flat^{d-2} 
    \left[ 1-{\frac{4(d-1)}{(d+1)}} {\frac{\eta_W}{L_\flat^2}} \right] \mathfrak{E}
+ {\frac{3}{4}}{\dot{V}}_\flat \Phi
\end{align}
where  
\begin{equation}
E(\rho) \equiv {\frac{Q_\flat}{L_\flat^4}}\mathfrak{E}(\rho) \,.
\end{equation}
Apart from dimensional analysis, all this simple redefinition does is to measure 
$E(\rho)$ perturbations in units of $Q_\flat$. This is a valid operation for this case of $AdS_2\times \mathbb{R}^d$, but would clearly make no sense in $AdS_D$ where $Q_\sharp=0$.
For the ${\rho\rho}$ gravity equation we get 
\begin{align}\label{ads2linpp}
& 0= - {\frac{4(d-1)}{(d+1)}}{\frac{\eta_W}{L_\flat^2}} (\rho^3 B''')
-{\frac{8(d-2)(d-1)}{(d+1)}} {\frac{\eta_W}{L_\flat^2}}  (\rho^2 B'')
\notag\\ & \quad
+\left[-d + {\frac{4d(d-1)}{(d+1)}} {\frac{\eta_W}{L_\flat^2}} \right] 
\left\{ (\rho B') - B \right\}
+ \left[ -d + {\frac{4(d-1)(d-2)}{(d+1)}}{\frac{\eta_W}{L_\flat^2}} \right] 
  (\rho^3 A')
\notag\\ & \quad
+ \left[ -(d-2) +{\frac{4(d-1)(d-6)}{(d+1)}} {\frac{\eta_W}{L_\flat^2}}\right] (\rho^2 A)
+ f_\flat L_\flat^{d-2} 
    \left[ 1-{\frac{4(d-1)}{(d+1)}} {\frac{\eta_W}{L_\flat^2}} \right] \mathfrak{E}
+ {\frac{3}{4}}{\dot{V}}_\flat \Phi
\end{align}
and for the $xx$ equation (identical to the other $x_i \, x_i$ equations) we get
\begin{align}\label{ads2linxx}
& 0= 
-{\frac{4(d-1)}{d(d+1)}}{\frac{\eta_W}{L_\flat^2}} (\rho^4 B'''')
+ \left[(d-1)+{\frac{8(d-1)^2}{d(d+1)}}{\frac{\eta_W}{L_\flat^2}} \right] 
\left\{ (\rho^2 B'') -2 (\rho B')\right\}
\notag\\ & \quad
+ \left[ d + {\frac{4(d-1)(d-2)}{d(d+1)}}{\frac{\eta_W}{L_\flat^2}} \right] 
\left\{  (\rho^4 A'') + 2 (\rho^2 A') \right\}
+ \left[ 2(d-1) + {\frac{16(d-1)^2}{d(d+1)}} {\frac{\eta_W}{L_\flat^2}}\right] B
\notag\\ & \quad
+ \left[-4 +{\frac{16(d-1)}{(d+1)}} {\frac{\eta_W}{L_\flat^2}}\right] (\rho^2 A)
+ f_\flat L_\flat^{d-2} 
    \left[ -1+{\frac{4(d-1)}{(d+1)}} {\frac{\eta_W}{L_\flat^2}} \right] \mathfrak{E}
+ {\frac{1}{4}}{\dot{V}}_\flat \Phi
\end{align}
The dilaton equation is again second order, 
\begin{equation}\label{ads2linP}
(\rho^2 \Phi'') 
- \left\{ (L_\flat^2{\ddot{V}}_\flat)  
    + {\frac{2{\ddot{f}}_\flat}{f_\flat}} 
  \left[ 1- {\frac{4(d-1)}{(d+1)}}{\frac{\eta_W}{L_\flat^2}} \right]
   \right\} \Phi 
+ \frac{2}{\rho^2} (L_\flat^{d} {\dot{V}}_\flat ) f_\flat \mathfrak{E} =0 \,,
\end{equation}
and the electric perturbation equation is
\begin{equation}\label{ads2linE}
\left( 1-{\frac{4\eta_W}{3L_\flat^2}} \right) f_\flat
\left[ -{\frac{1}{2}} \mathfrak{E}' + {\frac{1}{\rho}} \mathfrak{E} \right] 
- {\frac{1}{4}}(L_\flat^2 {\dot{V}}_\flat) \left[ \rho^2\Phi' - 2\rho\Phi \right] =0 \,.
\end{equation}

In similar spirit to the case of $AdS_D$ in the previous subsection, the form of the perturbation equations about $AdS_2\times \mathbb{R}^d$ permits {power law} solutions of the form
\begin{equation}
\Phi  = \breve{\Phi} \rho^{\nu_\Phi} \,, \qquad
A =  \breve{A} \rho^{\nu_A} \,,\qquad
B =  \breve{B} \rho^{\nu_B} \,,\qquad
\mathfrak{E} =  \breve{\mathfrak{E}} \rho^{\nu_E} \,,
\end{equation}
where $\{\breve{A},\breve{B},\breve{\Phi},\breve{\mathfrak{E}} \}$ are simple constants,
as long as the metric indices $\nu_A, \nu_B$ are tied to the dilaton index $\nu_\Phi$ and the electric index $\nu_E$ by
\begin{align}\label{ads2r2nuABE}
\nu_A & =\nu_B - 2 \,, \\
\nu_\Phi & = \nu_B- 1\,,\\
\nu_E & =\nu_B + 1 \,.
\end{align}
For the $\nu_\Phi$ index, we find
\begin{equation}
\nu_\Phi^\flat = {\frac{1}{2}} \pm 
\sqrt{ {\frac{1}{4}}+ (L_\flat^2{\ddot{V}}_\flat) - 
2{\frac{{\ddot{f}}_\flat}{f_\flat}} \left( 1-{\frac{4\eta_W}{3L_\flat^2}} \right) 
- 2 (L_\flat^2 {\dot{V}}_\flat) {\frac{\mathfrak{E}_\flat}{\Phi_\flat}}   
} \,.
\end{equation}

Note that $\nu_\Phi^\flat$ depends on three theory function derivatives: $(L_\flat^2{\ddot{V}}_\flat)$, $({\ddot{f}}_\flat/f_\flat)$, and $(L_\flat^2{\dot{V}}_\flat)$, as well as one solution parameter $\mathfrak{E}_\flat/\Phi_\flat$ (via the equation of motion) and the Weyl squared correction parameter $\eta_W$. In the previous subsection we already found the condition involving $L_\sharp^2{\ddot{V}}_\sharp$ required to keep $\Phi$ non-tachyonic in $AdS_D$. We now examine the two new pieces in order to see if our theory functions and theory parameters can support dilaton perturbations about $AdS_2\times \mathbb{R}^d$ with real values of $\nu_\Phi$. Note that such subtleties did not arise in the magnetic Lifshitz crossovers obtained in \cite{Kachru_Lif_2012} in a context without Weyl squared corrections to the gravity sector.

We need to know the sign and magnitude of $-2{\ddot{f}}_\flat/f_\flat$. Calculating it from the implicit form for $f(\phi)$ from Section (\ref{sec:r2hsv}) and going into the $AdS_2\times \mathbb{R}^d$ region, we find a number of order one regardless of $\beta$ or $\alpha$,
\begin{equation}
-2\left.{\frac{{\ddot{f}}_\flat}{f_\flat}}\right|_{AdS_2\times \mathbb{R}^d} =
 -{\frac{4}{D_3}} < 0 \,.
\end{equation}
Insisting that this term does not overwhelm the 1/4 under the square root gives a condition on solution parameters, 
\begin{equation}
d^{2} (z-1) + (16-d)\theta > 16 d \,.
\end{equation}
This condition is compatible with the conditions we found on the existence of UV crossovers to $AdS_{D}$: If $d=2$, we found that we needed $z=4$ for an $AdS_{D}$ crossover, in the IR case, then we need $\theta > 20/14$.  For $d>2$, we find a range of $z>1$ which falls within that required for the existence of UV $AdS_{D}$ crossovers.

The third piece in the surd is more opaque, and interesting because it can compete against the other two terms. We would like to ensure that $(L_\flat^2{\dot{V}}_\flat)$ has a consistent sign for all $\phi$ i.e. for all $r$. This is important because if $(L_\flat^2{\dot{V}}_\flat)$ could be zero at some $r$ (equivalently, at some $\phi$), then the $\mathfrak{E}_\flat/\Phi_\flat$ term could drive the $\nu_\Phi$ imaginary during the crossover evolution, which would not be physical. Calculating $L_\flat^2{\dot{V}}_\flat$, we find
\begin{equation}
-2 ( L_\flat^2{\dot{V}}_\flat )= + {\frac{2\sqrt{2}}{\sqrt{D_3}}} (\beta-1) D_1 >0 \,,
\end{equation}
because $\beta>1$ in the physical range of parameters. This holds for all parameters, so it is true as a theory function statement.  

So far, we have only explored curvature squared corrections of Weyl type, with dilaton-dependent theory function coefficients. It would be interesting to know if it is possible to find crossovers with dilaton-dependent theory function coefficients for Gauss-Bonnet and $R^2$ terms as well, although we have not investigated this.

%========+=========+=========+=========+=========+=========+=========+=========+
\section{Summary of findings and outlook}\label{sec:discussion}

In section 2, we find hyperscaling violating (HSV) solutions to an Einstein-Maxwell-dilaton model with curvature squared corrections (with constant coefficients) and $f(\phi)$ gauge coupling with dilaton potential $V(\phi)$. We make a simple isotropic, static, spherically symmetric ansatz for the HSV metric, and solve the equations for HSV solution parameters $z,\theta$ depending on theory parameters $d,\{\eta_i\}$. From a bottom-up perspective, insisting on having HSV solutions in this curvature squared model puts conditions on the functions $V(\phi),f(\phi)$. 

Generally, the expression for $\phi$ is a competition between a[n asymptotically] logarithmic piece ${\rm arccsch}(a\,r^{2\theta/(d-\theta)})$ and a power law piece, $\sqrt{1+ a^2\, r^{2\theta/(d-\theta)}}$, where $a$ is linear in $\{\eta_i\}$ and depends on $d,z,\theta$. From this, we can see immediately that turning off the $\{\eta_i\}$ removes the power law piece and leaves only the log, making $V(\phi)\sim- e^{-b\phi}$, where $b=b(z,\theta,d)$, which is the behaviour previously found with Einstein gravity \cite{Kachruhsv}.
With the curvature squared terms turned on, obviously the character of the equations of motion changes, and this alters the dependence on $\phi$ of the dilaton potential $V(\phi)$, in  such a way that we do not recover simple exponentials asymptotically far out: $V(\phi)\sim -c_1\phi^4-c_2\phi^2$, where $c_i=c_i(d,z,\theta,\{\eta_i\})$. This is a nuance of order of limits.  Interestingly, deep in the interior it reduces to a sum of exponentials: $V(\phi)\sim -e^{-2c\phi}-e^{-4c\phi}$, where $c=c(z,\theta,d)$. 
Now let us comment on the gauge coupling function $f(\phi)$.
In the limit that $\{\eta_i\}\rightarrow 0$, $f(\phi)\sim e^{-f_1\phi}$, where $f_1=f_1(z,\theta,d)$. Far out in the geometry, 
$f(\phi)\sim \phi^{-f_2}$, where $f_2=f_2(\theta,d)$, 
while deep in the interior $f(\phi)\sim \infty$, which forces $F^2$ to zero there.

In section 3 we shift gears to constraining allowed ranges of parameters for HSV solutions by using the null energy condition. The NEC  restricts polynomial combinations of solution parameters $z,\theta$ (or equivalently $\alpha,\beta$) and theory parameters $\{\eta_i\}$. These constraints look opaque at first, so we investigate them graphically in stages of complexity. We first visualize the HSV NEC constraints for single $\eta$ first, then pairs of $\eta_i$, then all three at once. 

{\underline{Weyl}}: For $d=1$ the $C^2$ term does not contribute and the NEC reduces to that of Einstein gravity, as studied in e.g. \cite{sachdevhsv}.  
For $d=2$,  for $\eta_W>0$, we find $1\leq z < 4$, while for $\eta_W<0$ $z>4$ . (Recall that we do not consider $z<1$ for causality reasons.) 
At $z=4$, $\eta_W$ is unconstrained: the Weyl term simply does not contribute to the equations of motion. 
For $d>2$, qualitatively similar behaviour ensues: the range of $z$ is now $\theta$-dependent. 
When $\eta_W$ is positive, we find small $z,\theta$. For negative $\eta_W$ we find $z>4$; which $\theta$ are admissible depends on $d$.

{\underline{Gauss-Bonnet}}: For $d=1$ and $d=2$, the $\eta_{GB}$ terms vanish from the equations of motion, as expected because the bulk Gauss-Bonnet action is identically zero for $d=1$ ($D=3$) and topological for $d=2$ ($D=4$). For $d>2$, for $\eta_{GB}<0$, the NEC restrictions are the same as for Einstein gravity, whereas for $\eta_{GB}>0$, only $z=1,\theta=0$ is admissible (plain $AdS$).

{\underline{$R^2$ gravity}}: For $\eta_R>0$, only $z=1,\theta=0$ is admissible.
For $\eta_R<0$, we find two distinct cases.
(a) For $z>4$, any $\theta$ is admissible. 
(b) For $z<4$, only some $\theta$ are allowed; the curved constraint surface
is quartic in $\theta$ and cubic in $z$. There are also two other physical constraints illustrated in the plots: the requirements that (i) $d_{\rm eff}\equiv d-\theta \geq 0$ and (ii) $\theta \geq 0$, which were motivated from the condensed matter side. They are visible in the plots as planar edges to permissible parameter ranges.

{\underline{$R^2$ and Weyl}}: For $R^2$ Weyl, ranges of permissible $z,\theta$ arise for all four sign choices of the $\eta_W,\eta_R$ parameters. The behaviour changes radically at $z=4$, for any $d$. 
(a) For $z=4$, we need $\eta_R<0$. For $d=2$ only, $\eta_W$ is unconstrained. For $d>2$, $\eta_W$ can be positive or negative depending on $\eta_R$. (For $d=1$, 
the Weyl tensor vanishes so there is no $\eta_W$.)
(b) For $z>4$, $\eta_R$ must still be negative; $\eta_W$ can be positive or negative depending on $\eta_R$.  
(c) For $z<4$, $\eta_R$ can be positive or negative and $\eta_W$ can be positive or negative, depending on $\theta$. Positive values of $\eta_R$ only occur for $\eta_W$ positive. Also, to have $\theta$ take every value between 0 and $d$ requires $\eta_W>0$, whereas for cases (a) and (b) $\theta$ can take any value within its physical range.

{\underline{$R^2$ and Gauss-Bonnet}}: In this case, by having $\eta_R$ turned on, we can now access values of $\eta_{GB}$ that were previously off limits. The behaviour of the plots again changes qualitatively at $z=4$ for any $d\geq 3$. (The cases $d=1,2$ are not discussed as the Gauss-Bonnet term vanishes, reducing the pair of parameters to a single one covered previously.)
(a) For $z<4$, both positive and negative $\eta_R$ and $\eta_{GB}$ are allowed up to a maximum value of $\theta$ less than $d$. 
(b) For $z\geq 4$, only $\eta_{GB}<0$ is allowed and $\eta_{R}$ sits within a restricted range that goes from just above zero to the negative region.

{\underline{Gauss-Bonnet and Weyl}}: For $d=1,2$ the Gauss-Bonnet term vanishes and we reduce back to the Weyl-only case. We therefore take $d\geq 3$.
The behaviour in this case depends on $z$.
(a) $z<4$:  All positive values of $\eta_W$ are allowed; not all negative values are allowed, the boundary of the range depending on $\theta$. $\eta_{GB}$ can be positive or negative depending on $\theta$ (as well as $d,z$, of course).
(b) $z>4$: Both $\eta_W$ and $\eta_{GB}$ can be positive or negative, with the permissible ranges depending on each other and on $\theta$ (as well as $d,z$, of course).

{\underline{$R^2$, Gauss-Bonnet, and Weyl}}: Having all three curvature squared theory parameters turned on is obviously the most complex case. Consider the full NEC conditions (\ref{D4RGBW}) and (\ref{D6RGBW}) on the constants $D_4$ and $D_6$.
First, we can examine the terms proportional to $\eta_{GB}$. These vanish identically in $d=1,2$. Since these terms have factors identical to the factors in the Einstein-only NEC, this implies that when there is only $\eta_{GB}$ it has to be negative; however, when they are turned on, we can access previously prohibited values of $\eta_{GB}$.
Second, consider the terms in $\eta_R$. The $D_6$ coefficient vanishes when $z=4$ in $d=2$, and can also vanish in $d>2$ but with $z$ now depending on $\theta$ in the physical range. Then the plots would just transition back to the two-parameter case studied in Section \ref{sec:GBW}. (See above for summary.)
The cleanest transition is seen in $d=2$ but it is morally similar in higher $d$. The $D_4$ coefficient, on the other hand, cannot vanish in the physical range of parameters.
Finally, consider the terms in $\eta_W$. The $D_4$ coefficient vanishes when $z=4$ in $d=2$; for the same values, $D_6$ is positive, implying that $\eta_W<0$. For other $z,d$, the constraints on $\eta_W$ are less severe: $\eta_W$ can take on positive or negative values depending on $\theta,z,d$.  

Generally, we see `features' in the plots of Section \ref{sec:plots} when particular terms in the the curvature squared NEC constraints vanish or change sign. This is most notable when the $\eta_R$ term is turned on, as its change in behaviour is the cleanest. In all the subsections except Gauss-Bonnet, we found that $z=4$ was a special value dividing different types of behaviours. The overall message is that physically acceptable HSV solutions are supported for a range of $\{\eta_i\}$ parameters, depending on $d$ and solution parameters $z,\theta$. 

In Section \ref{sec:crossover} we study the question of crossover solutions between $AdS_D$ near the boundary, HSV in the intermediate region, and $AdS_2\times \mathbb{R}^d$ in the deep interior. The idea is that the $AdS_D$ completion provides a resolution of the large-$r$ curvature singularity of the HSV space, while the $AdS_2\times \mathbb{R}^d$ completion provides a resolution to its tidal force singularity at small $r$. The Weyl corrected Einstein-Maxwell-dilaton theory does support both $AdS_D$ and $AdS_2\times \mathbb{R}^d$ solutions. By linearizing the equations of motion about those backgrounds, we see that perturbations taking us from either background to HSV require conditions on $z,\theta,d$. For $AdS_D$ the condition is 
\begin{equation}
d=2: \quad z=4 \,, {\rm any}\ \theta \in [0,d) \,,
\end{equation}
or 
\begin{equation}
d>2: \quad z<{\frac{2d}{d-1}} \quad {\rm for}\ \ 
\theta={\frac{d(2d+z-dz)}{d-2}} >0 \,.
\end{equation}
For $AdS_2\times \mathbb{R}^d$, we find an additional conditions on $z,\theta$,
\begin{equation}
d^2(z-1)+(16-d)\theta>16d
\end{equation}
It is possible to satisfy {\em both} conditions for a range of physically sensible $z,\theta$ of order one. 

A natural question to ask would be how to generalize our results to other systems with hyperscaling violation. We focused here on the electric ansatz; it would be interesting to know what might change with using instead a magnetic or (in $d=2$) even a dyonic ansatz. Another interesting direction would be to consider models with more complicated ansatze, such as those with Bianchi type symmetries in the boundary directions as in \cite{bianchi2}. It is also important to work out how such bottom-up constructions might mesh with supergravity/string embedding, as in  e.g. \cite{Haack}.

Entanglement entropy $S_{EE}$ is interesting because it provides a non-local probe in general AdS/CFT contexts which is different in character from Wilson loops. Calculating $S_{EE}$ from field theory is notoriously difficult, but when the gravity dual is described by Einstein gravity with matter it can be calculated holographically via the Ryu-Takayanagi (RT) formula \cite{HoloEE}. The RT formula calculates the area of the minimal surface which extends into the bulk and is homologous to the entangling region on the boundary. For HSV solutions in Einstein gravity, $S_{EE}$ displays logarithmic violations of the area law, as expected for condensed matter systems with Fermi surfaces \cite{sachdevhsv}.

For general $\{\eta_{GB},\eta_W, \eta_R\}$ curvature squared corrections, the formula for the entanglement entropy is not yet known (see \cite{Dong_2013} and \cite{Camps_2013} for recent progress along these lines), except for the sub-case where only $\eta_{GB}$ is turned on. Myers et al proposed a generalization of the Ryu-Takayanagi formula to Lovelock gravity \cite{entlove}.   
The entanglement entropy for our HSV solutions with $\eta_{GB}$ turned on 
can in principle be computed from the Myers et al $S_{EE}$ formula \cite{entlove}. 
In $d=2$, relevant for condensed matter, because the 
the $\eta_{GB}$ term is topological, it does not contribute in the equations of motion, so it should yield the same result as for Einstein gravity. 
It would be interesting to do the explicit (hard) computation using \cite{entlove} of the entanglement entropy in this case to check explicitly. Physically, the important question to resolve is whether or not there are log violations in $S_{EE}$ for HSV solutions in gravity theories with curvature squared corrections.  

%========+=========+=========+=========+=========+=========+=========+=========+
\section*{Acknowledgements}

The authors wish to thank Dr.~Benjamin A. Burrington and Dr.~Ida G. Zadeh for useful discussions and comments. 

DKO wishes to thank the Ludwig-Maximilians-Universit{\"{a}}t M{\"{u}}nchen for hospitality during the August 2013 Arnold Sommerfeld School on gauge-gravity duality and condensed matter physics. AWP wishes to thank the University of Michigan Center for Theoretical Physics for hospitality during the October 2013 Black Holes in String Theory workshop.

This research was supported by the Natural Sciences and Engineering Research Council (NSERC) of Canada. The research of DKO was supported in part by an E.F.~Burton Fellowship. 

%========+=========+=========+=========+=========+=========+=========+=========+
\appendix

\section{Appendix} \label{appendix}

In this appendix we provide, for completeness, the constants $C_{i}$ that arise in the equations of motion and in the NEC.  They are
\begin{equation}
C_{1}(d,\beta) = -\frac{d}{2}(d + 2\beta -1) \,,
\end{equation}
\begin{align}
C_{2}(d,\alpha,\beta,\eta_{i}) =  &-\frac{1}{2L^{2}} \left . \bigg{\{} \big{[}4 \alpha^{4} -8(1-\beta) \alpha^{3} +(-44 + 88 \beta -40 d \beta  - 44 \beta^{2} - 8d^{2} + 44 d) \alpha^{2} \right . \notag \\
&\left . + (-144 \beta  + 16 d^{2} -64 d +104 d \beta + 48 - 8 d^{2} \beta +144 \beta^{2} - 40d \beta^{2} -48 \beta^{3} ) \alpha \right . \notag \\
&\left . + 2d - 4 d \beta^{2} -2 d^{2} \big{]} \eta_{1} +\big{[}2 \alpha^{4} +(2d +4\beta -4) \alpha^{3}  \right . \notag \\
&\left . +(13d - d^{2} + 44\beta -22 -22 \beta^{2} -12 d \beta) \alpha^{2} \right . \notag \\
& \left . + (8 d^{2} -6 d^{2}\beta + 72 \beta^{2} +24 -72 \beta -32 d -24 \beta^{3} -26 d \beta^{2} + 58d \beta)  \alpha \right . \notag \\
&\left . +5d -12 \beta^{3} d -11 d^{2} \beta^{2} +16 d^{2} \beta +29 d \beta^{2} -6 d^{2} -22 d \beta + d^{3} -2 d^{2} \beta \big{]} \eta_{2} \right. \notag \\
&\left . +\big{[}(2\alpha \beta + 2d\beta - 2\alpha +2\alpha^{2} -d +2 d\alpha +d^{2} ) \times (-24 \beta^{2} + 48 \beta -24  \right . \notag \\
&\left . +2\alpha\beta-2\alpha + 2\alpha^{2} - 10d \beta + 9d +2 d\alpha -d^{2})\big{]} \eta_{3} \right . \bigg{\}}   \,,
\end{align}
\begin{equation}
C_{3}(d,\alpha) = \frac{d}{2}(d + 2\alpha - 1) \,,
\end{equation}
\begin{align}
C_{4}(d,\alpha,\beta,\eta_{i}) = \frac{1}{2L^{2}} &\left .  \bigg{\{} \big{[}-12d\beta^{2}-8\alpha^{3}\beta + 20\alpha^{2} d - 8 d \alpha \beta + 24 \alpha^{2} \beta - 12 \alpha^{2} \beta^{2} \right . \notag \\
&\left . + 8 \alpha^{3} - 12 \alpha^{2} + 6 d^{2} + 4 \alpha^{4} - 8 d \alpha^{3} - 8 d^{2} \beta + 24 d \beta - 6 d \big{]} \eta_{1} \right . \notag \\
&\left . +\big{[}13 \alpha^{2} d - 3 d + 4\alpha^{3} - 6 \alpha^{2} + 2 d^{2} + 2 \alpha^{4} + 4 d^{2} \beta - 4 \alpha^{3} \beta - \alpha^{2} d^{2} \right . \notag \\
&\left . + d^{3} - 6 \alpha \beta d^{2} + 10 d \alpha \beta + 6 d \beta -  6 \alpha^{2} \beta^{2} + 4 d^{2} \alpha - 4 d \alpha - 2 d \alpha^{3} \right . \notag \\
&\left . +12 \alpha^{2} \beta - 3 d \beta^{2} - 8 d \alpha^{2} \beta - 6 d \beta^{2} \alpha - 2 d^{3} \beta - 3d^{2} \beta^{2} \big{]} \eta_{2} \right . \notag \\
&\left . \big{[}2\alpha^{2} + 2 \alpha \beta - 2 \alpha + 2 d\alpha + 2d \beta + d^{2} -d) \right . \notag \\
&\left . \times (-6 d \beta - 6 \alpha \beta + 6 \alpha +2 \alpha^{2} + 9 d - 2 d \alpha - d^{2}\big{]} \eta_{3} \right . \bigg{\}} \,,
\end{align}
\begin{equation}
C_{5}(d,\alpha,\beta) = \alpha^{2} + (d + \beta - 2) \alpha + \frac{1}{2} d (d - 3) + (d-1)\beta + 1 \,,
\end{equation}
\begin{align}
C_{6}(d,\alpha,\beta,\eta_{i}) = &-\frac{1}{2L^{2}} \left . \bigg{\{} \big{[}28 \alpha^{2} - 16 \alpha^{3} - 24 \alpha^{2} \beta + 4 \alpha^{4} + 8 \alpha^{3} \beta + 8 \alpha \right . \notag \\
&\left .  - 4 d \alpha^{2} + 48 \beta^{3} + 30 d - 24 + 44 d \beta^{2} - 6 d^{2} + 4 \alpha^{2} \beta^{2} + 120 \beta + 16 d \alpha \beta \right . \notag \\
&\left . - 144 \beta^{2} + 40 \alpha \beta^{2} - 56 \alpha \beta + 8 d^{2} \beta - 80 d \beta - 8 d \alpha \big{]} \eta_{1} \right . \notag \\
&\left . + \big{[}-12 + 11 d + 8 \alpha^{3} \beta - 48 \alpha^{2} \beta + 18 \alpha^{2} \beta^{2} - 25 d \beta^{2} + 28 \alpha \beta  \right . \notag \\
&\left . + 12 d \alpha + 2 d \beta + 12 \beta^{3} \alpha +12 d \beta^{3} - 8 \alpha + 36 \beta + 2 \alpha^{4} - 12 \alpha^{3} + 30 \alpha^{2} \right . \notag \\
&\left . - 36 \beta^{2} + 12 \beta^{3} + 2 d^{2} - 32 \alpha \beta^{2} -13 d \alpha^{2} - 4 d^{2} \alpha - 16 d^{2} \beta + 2 d \alpha^{3} \right . \notag \\
&\left . + d^{2} \alpha^{2} + 11 d^{2} \beta^{2} + 2 d^{3} \beta - 38 d \alpha \beta + 12 d \alpha^{2} \beta + 22 d \beta^{2} \right . \notag \\
&\left . + 6 d^{2} \alpha \beta - d^{3} \big{]} \eta_{2} +\big{[}(2 \alpha^{2} + 2 \alpha \beta - 2 \alpha + 2 d \alpha + 2 d\beta + d^{2} - d) \times (24 \beta^{2}  \right .\notag \\
&\left . - 60 \beta  + 36 + 10 \alpha \beta - 14 \alpha + 2 \alpha^{2} + 10 d \beta - 13 d + 2 d \alpha + d^{2}) \big{]} \eta_{3} \bigg{\}} \right. \,. 
\end{align}

%========+=========+=========+=========+=========+=========+=========+=========+

\end{document}